\documentclass[article,aps,pra,twocolumn,showpacs,superscriptaddress,groupedaddress]{revtex4}  
\usepackage{graphicx}  
\usepackage{dcolumn}   
\usepackage{bm}        
\usepackage{amssymb}   
\usepackage{amsmath}
\usepackage{braket}
\usepackage[caption=false]{subfig}
\usepackage{verbatim}
\usepackage{color}
\newcommand{\be}{\begin{equation}}
\newcommand{\ee}{\end{equation}}
\newcommand{\ben}{\begin{eqnarray}}
\newcommand{\een}{\end{eqnarray}}
\newcommand{\bes}{\begin{subequations}}
\newcommand{\ees}{\end{subequations}}
\newcommand{\bF}{\begin{figure}}
\newcommand{\eF}{\end{figure}}

\newcommand{\RNum}[1]{\uppercase\expandafter{\romannumeral #1\relax}}

\def\tr{ {\rm{Tr }}\,}

\newcommand{\kt}{\rangle}
\newcommand{\br}{\langle}

\newcommand\numberthis{\addtocounter{equation}{1}\tag{\theequation}}
\def\tr{ {\rm{Tr }}\,}

\begin{document}
\title{Out-of-time-ordered correlators and the Loschmidt echo in the quantum kicked top: How low can we go?}

\author{Sreeram PG}
\affiliation{Department of Physics, Indian Institute of Technology Madras, Chennai, India 600036}

\author{Vaibhav Madhok}
\affiliation{Department of Physics, Indian Institute of Technology Madras, Chennai, India 600036}

\author{Arul Lakshminarayan}
\affiliation{Department of Physics, Indian Institute of Technology Madras, Chennai, India 600036}

\begin{abstract}
The out-of-time-ordered correlators (OTOC) and the Loschmidt echo are two measures that are now widely being explored to characterize sensitivity to perturbations and information scrambling in complex quantum systems. Studying few qubits systems collectively modelled as a kicked top, we solve exactly the three- and four- qubit cases, giving analytical results for the OTOC and the Loschmidt echo. While we may not expect such few-body systems to display semiclassical features, we find that there are clear signatures of the exponential growth of OTOC even in systems with as low as 4 qubits in appropriate regimes, paving way for possible experimental measurements. We explain qualitatively how classical phase space structures like fixed points and periodic orbits have an influence on these quantities and how our results compare to the large-spin kicked top model. Finally we point to a peculiar case at the border of quantum-classical correspondence which is solvable for any number of qubits and yet has signatures of exponential sensitivity in a rudimentary form. 
\end{abstract}
\maketitle

\section{Introduction}{\tiny}

The contemporary interest and progress in quantum information processing have happened along with control over single
or few particle systems that are driving home the novelty of unique quantum phenomena such as
entanglement. It has also opened doors for investigation in the time domain, with exquisite control of individual quantum systems in the laboratory and the ability to drive these systems with designer Hamiltonians that can simulate phenomena as diverse as many-body-localization to ergodicity, chaos 
and thermalization. Two experiments that preserve the coherence and purity of complex many-body
time-evolving states illustrate the richness of this domain \cite{Neill16, Kaufman2016}. 

The first of these \cite{Neill16} involved the study of 3 qubits in a superconducting transmon setup that simulated the quantum kicked
top. Using state tomography they made connections between the onset of chaos and concomitant enhancement
in the entanglement. The second \cite{Kaufman2016} involved a two-dimensional Bose-Einstein condensate of $^{87}$Rb atoms,
implementing effectively a 6 particle Bose-Hubbard Hamiltonian. The study of thermalization via the development
of entanglement in such experiments on isolated quantum systems is of interest in the foundations of statistical mechanics,
and they test the Ergodic Thermalization Hypothesis (ETH) that is currently of great theoretical interest as well. 
Connections between low-dimensional ergodicity and chaos with entanglement, general quantum correlations and state tomography have long been studied, mostly theoretically, (for example in \cite{MillerSarkar,Lakshminarayan,BandyopadhyayArul2002,Tanaka-2002,LakSub2003,Bandyopadhyay04,Ghose2004,ArulSub2005,trail2008entanglement,LombardiMatzkin2011,mrgi14, madhok2015signatures, Madhok2018_corr,Maciej-2019,Meenu-2019,Pappalardi-2020}), although a cold-atom experiment as early as 2009 \cite{Chaudhary} was a pioneering work in this direction. 

These experiments also beg the question of how statistical properties such as thermalization and semiclassical 
properties such as chaos manifest in such low-dimensional quantum systems. The 3-qubit transmon experiment is based on 
the mapping of the well-studied quantum kicked top to a many-spin Floquet system. However, while traditional
studies of quantum chaos are for large spin $j$ \cite{Haake}, this experiment involved only $j=3/2$ and the mapped
system is in fact a nearest neighbor transverse field Ising model which is integrable. In any case, the solvability 
of this as well as the $j=2$ system which involves non-integrable next-nearest-neighbor interactions was demonstrated
in \cite{dogra19}. Such a study did show that it is possible to see some generic features and even some random matrix theory 
properties in such small systems. For example, it showed how with increasing the parameter controlling the non-integrability, entanglement moves
from being bipartite to multipartite, sharing it globally and demonstrating its monogamous nature.

Starting from \cite{RuebeckArjendu2017} which considered just two qubits ($j=1$) analytically and 3 qubits ($j=3/2$) numerically,
there have been studies that followed the fate of the few qubit kicked top \cite{dogra19,Madhok2018_corr,Bhosale-2018}. 
A recent experiment \cite{MaheshUdayExpt-2019} used NMR to study the 2 qubit version of the kicked top already displaying some semiclassical features but 
also peculiar quantum ones such as time- and parameter- periodicity \cite{Bhosale-2018}.

However, most of the traditional signatures of quantum chaos are based on statistical spectral properties, such as
nearest-neighbor spacing statistics \cite{Haake}, which do not make sense for small systems. As a matter
of principle, the effective Planck constant $h_{\text{eff}}$ in such systems is large and hence quantum-classical correspondence time scales, such as
the Ehrenfest time of $E_f\sim \log(1/h_{\text{eff}})/\lambda_C$, where $\lambda_C$ is the classical Lyapunov exponent
are very short. Quantum properties such as superposition manifest rapidly in time and thus experiments that involve coherent time-evolution also seem to be out of reach of semiclassical features.

The present work is placed in this context as one that explores how two measures based on the time-evolution fare in ferreting out
non-integrability and chaos out of small quantum systems that are already experimentally realizable, hence the question is how low can we go? These measures are the out-of-time-ordered correlator (OTOC), being intensely studied now in a remarkable variety of contexts, and the Loschmidt echo, which has a longer history of study in low-dimensional chaos. We find that although only very short-time information is available, OTOC of $j=2$ and $j=5/2$ kicked-tops already show definite precursors of exponential growth, and many properties of the echo are also shared by large $j$ systems, although the exponential decay may not be apparent, at least in the regimes we have addressed here. Thus the answer to the question seems to be ``pretty low".
We also initiate the study of a kicked top of arbitrary spin $j$, but when the chaos parameter is so ``absurdly large" that the Lyapunov exponent $\lambda_C$ is as large as $\log(1/h_{\text{eff}})$, and the Ehrenfest time is still of order 1! This ``dual" case also manifests for low values of $j$, one does not require is a very large value of the chaoticity parameter for the top. The kicked top Floquet operator, as we shall discuss, can be written as a {\it sum} of just $4$ rotations (for integer $j$), and hence the interactions need not be implemented at all.

An array of quantum signatures of chaos have already been studied. Fidelity decay in quantum systems \cite{per00, sc96}, level statistics \cite{Berry375, bohigas1971spacing}, properties of regular and irregular wavefunctions \cite{Berry77a, BERRY197926, voros1976semi} and quantum scars  \cite{heller1984bound}, signatures in single particle billiards \cite{McDonald, robnik1985classical}, semiclassical trace formulas \cite{gutzwiller1971periodic} and imprints on quantum correlations and tomography  \cite{ArulSub2005,Bandyopadhyay04,BandyopadhyayArul2002,Chaudhary,Ghose2004,LakSub2003,Lakshminarayan,LombardiMatzkin2011, madhok2015signatures, Madhok2018_corr, trail2008entanglement, MillerSarkar, mrgi14}.
Recent trends that focus on many-body systems,  include studies involving connections of quantum chaos to OTOCs, entropic uncertainty relations, and the rate of scrambling of quantum information in many-body systems with consequences ranging from the foundations of quantum statistical mechanics, quantum phase transitions, and thermalization on the one hand to the scrambling of information in many-body systems and black holes on the other hand \cite{swingle2016measuring, hayden2007black, MaldacenaSYK15, Maldacena2016, hartman2013time, shenker2014black, sekino2008fast, Lashkari2013, hosur2016chaos, IyodaSagawa}.

The OTOC, in their simplest form, captures the growth of the incompatibility between two operators, when one of them is evolved in the Heisenberg picture while the other is stationary \cite{Larkin-1969,swingle2016measuring, Maldacena2016,Cotler-2018,Hashimoto-2017,Swingle-2018,Carlos-2019}. From the commutator-Poisson bracket connection,
this gives an analog of the classical separation of two trajectories with quantum mechanical operators replacing the classical phase space trajectory. 
For two Hermitian observables, OTOC is given by
\begin{equation}\label{eqq}
C_{W,V}(\tau)=-\langle[W(x, \tau), V(y, 0)]^2\rangle, 
\end{equation}
where the local operators $W$ and $V$  act on sites $x$ and $y$ respectively and $W(x, \tau)=U^{\dagger}(\tau)W(x, 0)U(\tau)$ is the Heisenberg evolution of operator $W$ under unitary dynamics $U(\tau)$. The expectation value is taken with respect to the thermal state at some temperature which we take to be infinite. In sufficiently chaotic systems, the OTOC  essentially vanishes till the information of the operator perturbation at $x$ reaches $y$, during which phase the operator becomes highly nonlocal, an occurrence that is dubbed operator scrambling. Thereafter there is a rapid increase of the OTOC before
it saturates in a finite system at which stage the localized information at $x$ is considered to have been scrambled throughout the system, and 
it is not possible to recover it from any local subset. If the rapid increase of the OTOC is exponential $\sim e^{2 \lambda_Qt}$, $\lambda_Q$ has
been referred to as a quantum Lyapunov exponent. 

If the system has a bound spectrum this implies instability in a finite space and can be taken as a 
definition of quantum chaos. Thinking of systems with a well-defined semiclassical limit, note that simple systems such as the inverted parabolic potential $-x^2$ have trivially exponentially growing OTOC, but are of course not chaotic, but merely unstable. Similarly, there could be naturally isolated unstable orbits in an otherwise integrable system and special operators may still show exponential OTOC growth. Still, the jury is out on the role of OTOC in general and hence studying them in as many scenarios is of interest. Systems with well-defined semiclassical or classical limits are of special interest as it is well understood in what sense they are non-integrable and what the classical Lyapunov exponents are, and there have been several studies on this \cite{Rozenbaum17,Arul-Baker-2018,Saraceno-2018,Moudgalya-2018,Klaus-2018,Jalabert-2018,Ravi-Arul-2020}, including some on the quantum kicked top \cite{Akshay2018,sieberer2019digital,yin2020quantum}

Though quantum systems do not show sensitivity to perturbations in initial state vectors, integrable and chaotic quantum systems show remarkably different behavior and sensitivity when the system dynamics itself is perturbed \cite{peres2006quantum, peres1984stability}. One of the concepts used to capture this notion of quantum chaos is the Loschmidt echo that is related to the fidelity between the evolution of a quantum system with exact dynamics and propagation under a slightly perturbed Hamiltonian \cite{prosen2003theory, gorin2006dynamics, PhysRevE.68.036216}. Alternatively, this quantifies the distance between the forward propagation of a system and its time-reversed dynamics under small perturbations.
This is interesting as the question of time-reversal itself and its connections to chaos, both quantum and classical, has been one of the foundational questions in physics.  The debate around the microscopic origins of the second law of thermodynamics from underlying time-reversal invariant classical mechanics leads to interesting paradoxes. For example, could one reverse the momenta of all particles in a system causing the entropy to decrease thereby violating the second law \cite{peres2006quantum}? 
In this work, our focus is to study Loschmidt echo for a few qubit kicked top that is exactly solvable \cite{dogra19}.

Loschimidt echo is defined as 
\begin{equation}
F(t)=|\langle \phi  |e^{i H't} e^{-i Ht} |\phi \rangle|^2,
\end{equation}
where $H$ is the Hamiltonian for the forward evolution and $H'$ is the perturbed Hamiltonian representing
imperfect time reversal, {\it i.e.}, the Hamiltonian responsible for backward evolution. The perturbed evolution can be 
due to environmental noise and thus there is an intimate connection between Loschmidt echo and decoherence \cite{Zurek/Paz}
The Loschmidt echo has a rather complex behavior that depends on the state $|\phi\kt$, the nature of the Hamiltonian $H$--whether
it is integrable or not, the degree of chaos if it is not integrable and also on the strength of the perturbation that defines $H'$. In certain regimes,
an exponential decay of the fidelity has been observed with a rate that is the classical Lyapunov exponent.

Recently the question of sensitivity to perturbations is connected to the accuracy and robustness of quantum information processing devices. After all, the quantum device/simulator is a many-body complex quantum system and one needs to benchmark its accuracy \cite{sieberer2019digital, PhysRevE.62.3504, PhysRevE.62.6366}.
How does one trust a quantum simulator that invariably involves a many-body chaotic Hamiltonian with a rapid proliferation of errors, especially near a quantum critical point that is typically characterized by high entanglement/complexity and a large Schmidt rank of the system density matrix \cite{hauke2012can, georgescu2014quantum, deutsch2020harnessing}? While these questions have been under active research for many decades, only recently experiments have reached the level of sophistication and
control where non-integrability, chaos, and thermalization of closed quantum systems are studied by
manipulating individual interacting quantum bits. Another interesting avenue on the applications of Loschmidt echo is the application to quantum-limited metrology and making sensors. Since chaotic systems are sensitive to perturbations, this suggests a way to for high precision metrology \cite{fiderer2018quantum}. 

This paper is organized as follows. In Sec.~(\ref{sec:KTinto}) the kicked top is introduced and some of its classical properties are mentioned. A complete solution of the quantum problem for $3$ and $4$ qubit cases is also carried out, in the sense that explicit expressions for the powers of the Floquet operator are given in terms of the Chebyshev polynomials. In Sec.~(\ref{sec:OTOC})
the OTOC is derived for the 3 and 4 qubit kicked tops and their dependence on time and the chaoticity parameter is discussed. The OTOC 
is also compared with that for a larger number of spins, found numerically. The peculiar case when the number of spins is arbitrary by the chaos parameter is very large is also discussed in this section. {In Sec.~(\ref{sec:Echo}), the Loschmidt echo is discussed and we summarise and discuss future directions in  Sec.~(\ref{sec:Conc}).}

\section{The case of kicked top}
\label{sec:KTinto}
The quantum kicked top is  characterized by the angular momentum vector $(J_x,J_y,J_z)$, and the Hamiltonian\cite{KusScharfHaake1987,Haake,Peres02}  is given by
\begin{equation}
\label{Eq:QKT}
H=\frac{\kappa_0}{2j}{J_z}^2 \sum_{n = -\infty}^{ \infty} \delta(t-n\tau)+\frac{p}{\tau} \, {J_y}.
\end{equation}
It consists of rotations and impulsive rotations caused by periodic kicks at regular intervals of time $\tau.$ The time evolution of the top is given by the unitary operator
\begin{equation}
\mathcal{U} = \exp\left [-i (\kappa_0/2j \hbar) J_z^2 \right]\exp\left[-i (p/\hbar) J_y\right],
\end{equation}
which describes the evolution from one kick to the next. Angle of rotation about the $y$ axis is given by $p$, and $\kappa_0$ is the chaoticity parameter, which is a measure of the twist applied between kicks.
 Here  we set $\hbar=1$ and $p=\pi/2$.
In the limit of very large angular momentum, the classical limit is reached. $i^{th}$ iteration the classical map   of the unit 
sphere phase space $X_{i}^2+Y_{i}^2+Z_{i}^2=1$ onto itself 
is given by 
\begin{eqnarray}
X_{i}&=&Z_{i-1}\cos(\kappa_0 X_{i-1})+Y_{i-1}\sin (\kappa_0 X_{i-1}),\nonumber \\
Y_{i}&=&-Z_{i-1}\sin(\kappa_0 X_{i-1})+Y_{i-1}\cos (\kappa_0 X_{i-1}),\nonumber \\
Z_{i}&=&-X_{i-1}.
\end{eqnarray}
where $X_{i},Y_{i},Z_{i}=J_{x,y,z}/j$ 
\begin{figure}
	\centering
	\includegraphics[scale=0.65]{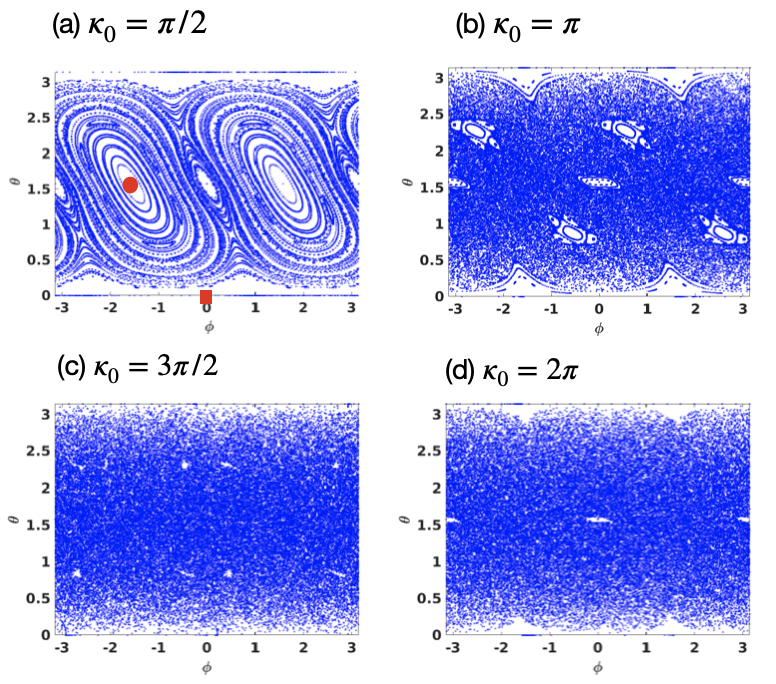}
	\caption{(a) Regular  (b,c)  mixed phase space and (d) chaotic phase space 
		resulting from the classical kicked top dynamics. Points labeled 
		with red square  and red circle in a) correspond to initial states $\Theta=0, \Phi=0$
		on a period-4 orbit
		and  $\Theta=\pi/2, \Phi=-\pi/2$ at the center of regular island respectively.}
	\label{fig:classical}
\end{figure}

Dynamics of a particle under these equations are simulated 
numerically for different initial states:
$(X_{0}, Y_{0}, Z_{0})$, and for two 
strengths of the chaos, $\kappa_0=0.5$ and $2.5$, as shown in Fig.~\ref{fig:classical},
conventionally termed as regular and mixed phase space structures respectively.
For $\kappa_0=0$ the classical map is integrable, being just a rotation, but for $\kappa_0>0$ chaotic orbits appear in the phase space, and when  $\kappa_0>6$ it is essentially fully chaotic. Connection to a many-body model can be made by considering the large $\mathbf{J}$ spin as the total spin 
of spin=1/2 qubits, replacing $J_{x,y,z}$ with $\sum_{l=1}^{2j} \sigma^{x,y,z}_l/2$ \cite{Milburn99,Wang2004}. The Floquet operator is then that of $2j$ qubits, an Ising model with all-to-all homogeneous coupling and a transverse magnetic field:
\begin{equation}
\label{uni}
{\mathcal U}=\exp\left(-i \frac{\kappa_0}{4j}  \sum_{ l< l'=1}^{2j} \sigma^z_{l} \sigma^z_{l'}\right)
\exp\left( -i \frac{\pi}{4} \sum_{l=1}^{2j}\sigma^y_l \right).
\end{equation}
Here $\sigma^{x,y,z}_l$ are the standard Pauli matrices, and an overall phase is neglected. The case of 2-qubits, $j=1$, has been analyzed in \cite{RuebeckArjendu2017} wherein interesting arguments have been proposed for the observation of structures not linked to the classical limit. In this case, several quantum correlation measures were also calculated in \cite{Bhosale-2018}.
For $j=3/2$, the three-qubit case is a nearest neighbor kicked transverse Ising model, known to be integrable \cite{Prosen2000,ArulSub2005}. For higher values of the spin, the model maybe considered few-body realizations of non-integrable systems. In general only the $2j+1$ dimensional permutation symmetric subspace of the full $2^{2j}$ dimensional space is relevant to the kicked top.

\subsection{Solving the the 3 and 4 qubit kicked tops}


The solutions in these cases were discussed first in \cite{dogra19},
where a wide variety of entanglement measures, from entropy to concurrence were studied and compared with
available experimental data. We recount here the essential details of the solutions for the sake of a self-contained narrative.
First, there is the general observation of an ``up-down" or parity symmetry: 
\[ [\mathcal{U},\otimes_{l=1}^{2j} \sigma^y_l]=0, \]
valid for any number of qubits. It is therefore optimal to work with a basis that is both permutation 
symmetric and is adapted to the parity.

For $j=3/2$ or the 3-qubit case, the standard 4-dimensional spin-quartet permutation symmetric space 
$\{|000\kt, |W\kt=(|001\kt+|010\kt+|100\kt)/\sqrt{3},
|\overline{W}\kt =(|110\kt+|101\kt+|011\kt)/\sqrt{3},|111\kt\}$
is parity symmetry adapted to form the basis
\begin{eqnarray}
|\phi^{\pm}_1\kt&=&\frac{1}{\sqrt{2}}(|000\kt \mp i | 111 \kt), \label{eq:3qubit_basis_1} \\
|\phi_2^{\pm}\kt&=&\frac{1}{\sqrt{2}} (|W\rangle \pm i |\overline{W}\kt).  \label{eq:3qubit_basis}
\end{eqnarray}
In this basis the Floquet unitary operator is given by
\begin{equation} 
\label{eq6}
\mathcal{U} = \begin{pmatrix}
\mathcal{U}_{+} & 0 \\ 0 & \mathcal{U}_{-}
\end{pmatrix}, 
\end{equation}
where $0$ is the $2 \times 2$ null matrix,
and $2\times2$-dimensional blocks $\mathcal{U}_{+}$ and $\mathcal{U}_{-}$ are written 
in the bases $\{ \phi_{1}^{+}, \phi_{2}^{+} \}$ and ($\{ \phi_{1}^{-}, \phi_{2}^{-} \}$),
and are referred to as positive and negative)-parity subspaces in our 
discussion. We have,
\begin{equation}
\mathcal{U}_{\pm} = \pm  e^{\mp \frac{i \pi}{4}} e^{-i \kappa} \begin{pmatrix}
\frac{i}{2}e^{-2i \kappa} & \mp \frac{\sqrt{3} }{2} e^{-2i \kappa} \\
\pm \frac{\sqrt{3}}{2} e^{2i \kappa} &  -\frac{i}{2}e^{2i \kappa}
\end{pmatrix},
\label{eq:Uplusm}
\end{equation}
corresponding to parity eigenvalue $\pm1$.
For simplicity the parameter $\kappa=\kappa_0/6$ is used in these expressions. 

Expressing Eq.~(\ref{eq:Uplusm}) as a rotation ($e^{-i \theta \sigma^{\hat{\eta}}}$)
by angle `$\theta$' about an arbitrary axis 
($\hat{\eta}=\sin{\alpha} \cos{\beta} \hat{x}
+ \sin{\alpha} \sin{\beta} \hat{y} + \cos{\alpha} \hat{z}$),
and a phase, we obtain,
$ \cos{\theta} =\frac{1}{2} \sin{2\kappa}$, 
$\beta=\pi/2 +2 \kappa$, and 
$\sin{\alpha} = \sqrt{3}/(2\sin{\theta})$.
Thus the time evolution is the propagator which is simply the power $\mathcal{U}^n$
is block-diagonal with blocks $\mathcal{U}_{\pm}^n$, which are 
explicitly given by,
\begin{equation}
\label{eq:Upluspowern}
\mathcal{U}_{\pm}^n = (\pm 1)^n e^{-i n (\pm  \frac{\pi}{4}+\kappa)}
\begin{pmatrix}
\alpha_n &
\mp \beta_n^* \\
\pm \beta_n &  
\alpha_n^*
\end{pmatrix}, 
\end{equation}   
where,
\begin{subequations}
\begin{align}
\label{eq:alpha3qub}
\alpha_n &= T_n(\chi)+\frac{i}{2}\, U_{n-1}(\chi) \cos 2\kappa \quad\\
\label{eq:beta3qub}
\beta_{n} &= (\sqrt{3}/2)\, U_{n-1}(\chi) \,e^{2i \kappa}.
\end{align}
\end{subequations}
Here the Chebyshev polynomials $T_n(\chi)$ and $U_{n-1}(\chi)$ of the first and second kinds are used and are defined as 
\begin{equation}
T_n(\chi)=\cos(n \theta)\;\; U_{n-1}(\chi)=\sin(n \theta)/\sin \theta,
\end{equation}
with 
$\chi=\cos{\theta}=\sin(2\kappa)/2=\sin(\kappa_0/3)/2$. Hence the matrix elements of the time $n$ propagator are explicitly given 
by polynomials of order $n$ in the variable $\sin(\kappa_0/3)$.

We further present an exact solution to a kicked top with spin $j=2$, modelled using four qubits, where each qubit is coupled to every other qubit by the same strength. Hamiltonian for such a system can be easily obtained from Eq.~\ref{Eq:QKT},
by substituting $j=2$. It is particularly interesting to study a four-qubit
kicked top as this is the smallest system where all-to-all interaction among qubits 
is different from that of nearest-neighbor interaction. 
Similar to that of the three-qubit kicked top, we are again confined to 
($2j+1$)-dimensional permutation symmetric subspace of the total $2^{2j}$-dimensional 
Hilbert space.
In this case the parity symmetry reduced and permutation symmetric basis
in which $\mathcal{U}$ is block-diagonal is
\begin{equation}
\label{eq:4QubBasis}
\begin{split}
|\phi_1^{\pm} \rangle &= \frac{1}{\sqrt{2}} (|W\rangle \mp | \overline{W} \rangle),\\
|\phi_2^{\pm} \rangle &= \frac{1}{\sqrt{2}} (|0000\rangle \pm | 1111 \rangle),\\
|\phi_3^{+} \rangle &= \frac{1}{\sqrt{6}} \sum_{\mathcal{P}}|0011\rangle_{\mathcal{P}}
\end{split}
\end{equation}
where $|W\kt =\frac{1}{2}\sum_{\mathcal{P}}|0001\kt_{\mathcal{P}}$, $|\overline{W}\kt =\frac{1}{2}\sum_{\mathcal{P}}|1110\kt_{\mathcal{P}}$, and $\sum_{\mathcal{P}}$ sums over all possible permutations. 

A peculiarity of 4-qubits is that $|\phi_1^{+}\kt$ is an eigenstate of $\mathcal{U}$ with eigenvalue $-1$ for {\it all} values of the parameter $\kappa_0$. Thus the $5-$ dimensional space splits into $1\oplus2\oplus2$ subspaces on which the operators are $\mathcal{U}_0=-1$ and $\mathcal{U}_{\pm}$. 
In this basis, the Floquet unitary operator $\mathcal{U}$ becomes block diagonal, 
which makes it easy to write the $n^{th}$ power of the
unitary operator $ \mathcal{U}$ as

\begin{equation}
\label{eq:4qubitUpown}
\mathcal{U}^n = \begin{pmatrix}
\mathcal{U}_0^n & 0_{1\times 2}& 0_{1\times 2} \\ 0_{2\times 1} &   
\mathcal{U}_{+}^n & 0_{2\times 2} \\ 0_{2\times 1} & 0_{2\times 2} &  \mathcal{U}_{-}^n
\end{pmatrix}, 
\end{equation}
This simplifies the problem significantly. 
Various blocks are written here explicitly, we have
\begin{equation}
\mathcal{U}_0=\langle \phi_1^{+} |\mathcal{U}|\phi_1^{+} \rangle = -1,
\end{equation}
which is a part of the positive-parity subspace.
Block $\mathcal{U}_{+}$ written in the $\{\phi_2^{+},\phi_3^{+}\}$ basis, is
\begin{equation}
\label{eq12}
\mathcal{U}_{+} = -ie^{-\frac{i \kappa }{2}} \left(
\begin{array}{cc}
\frac{i}{2} e^{-i \kappa} & \frac{\sqrt{3}i}{2}  e^{-i \kappa} \\
\frac{\sqrt{3}i}{2}  e^{i \kappa} & -\frac{i}{2} e^{i \kappa} \\
\end{array}
\right),
\end{equation}
while $\mathcal{U}_{-}$ in the basis $\{\phi_1^{-},\phi_2^{-}\}$, is
\begin{equation}
\label{eq12}
\mathcal{U}_{-} = e^{-\frac{3 i \kappa }{4}} \left(
\begin{array}{cc}
0 & e^{\frac{3 i \kappa }{4}} \\
-e^{-\frac{3 i \kappa }{4}} & 0 \\
\end{array}
\right),
\end{equation}
where $\kappa=\kappa_0/2$.

 In a manner similar to the case of 3-qubits above, the time $n$ propagator is now
 written compactly in terms of the Chebyshev polynomials. We have 
\begin{eqnarray}
\label{eq:4QubUppown}
\mathcal{U}_{+}^n &=&  e^{-\frac{i n(\pi+\kappa) }{2}} 
\begin{pmatrix}
\alpha_n & i\beta_n^{*} \\  i\beta_n & \alpha_n^{*}
\end{pmatrix},
\end{eqnarray}
where
\begin{subequations}
\begin{align}
\label{eq:4QubitAlphaBeta}
\alpha_n =&  T_{n}(\chi)+\frac{i}{2}U_{n-1}(\chi)\cos{\kappa} \\
\label{eq:beta4qub}
\beta_n  =&  \frac{\sqrt{3}}{2}U_{n-1}(\chi)e^{i\kappa},
\end{align}
\end{subequations}
with $\chi=\sin{\kappa}/2=\sin(\kappa_0/2)/2$.
The negative parity subspace evolution operator is
\begin{equation}
\label{eq:4QubUmpown}
\mathcal{U}_{-}^n = e^{-\frac{ 3in \kappa }{4}} \left(
\begin{array}{cc}
\cos \frac{n\pi}{2} & e^{\frac{3 i \kappa}{4}} \sin \frac{n\pi}{2} \\
-e^{-\frac{3 i \kappa }{4}} \sin \frac{n\pi}{2}  &  \cos \frac{n\pi}{2} \\
\end{array}
\right).
\end{equation}
Although for simplicity we use the same symbols $\alpha_n$ and $\beta_n$ for the propagator matrix entries in the 3 and 4 qubit
cases, they are not the same. However, in either case, we note the important identity that $|\alpha_n|^2+|\beta_n|^2=1$, following
 from the unitarity of the propagators involved, arises from the Pell equation for the Chebyshev polynomials:
\begin{equation}
\label{eq:Pell}
T_n(x)^2+(1-x^2) U_{n-1}^2=1.
\end{equation}

\section{OTOC and the kicked top}
\label{sec:OTOC}

The out-of-time-ordered correlators (OTOC) are closely connected to the growth 
of the incompatibility of observables due to the dynamics. They are currently being studied in a wide variety of contexts from many-body physics to field theories,
quantum gravity, and black holes in a remarkable coming together of many research 
communities. They are thought of as a way to investigate the ``quantum butterfly effect",
which was also the role and motivation for the introduction of the Loschmidt echo. 
Both of these quantities, in systems with a semiclassical limit, have 
regimes where the classical Lyapunov exponent plays a role: as (half) the rate of the exponential growth of OTOC and 
as the rate of exponential decay of the echo. The Lyapunov exponent may be seen more clearly
in the OTOC as the echo has a rather complex dependence on the perturbation used, however recent 
works have pointed out explicit connections between OTOC in an averaged sense and the echo \cite{Yan-2019}.

Let $A(0)$ be some observable and let $A(t)=\mathcal{U}^{-t} A(0) \mathcal{U}^t$ be its Heisenberg time evolution.  
We define the OTOC as
\begin{equation}
\label{eq:OTOCdefn}
C_{\rho}(t)=-\frac{1}{2}\tr \left(\rho\, [A(t),A(0)]^2 \right).
\end{equation}
where $\rho$ is some state of the system. In particular we deal with the infinite temperature state
$\rho=I/(2j+1)$ denote the corresponding OTOC as $C_{\infty}(t)$. 
The phrase ``out-of-time-ordered" is justified for these quantities as the commutator contains 
terms such as $\br A(t) A(0) A(t) A(0) \kt$ wherein the operators are not monotonically ordered in time.
OTOC have been used an indicator of information scrambling as some initially localized operator or 
``information" in a  many-body system gets entangled with other one-particle operators on other sites and 
leads to a complex state wherein the initial information is practically lost. For nonintegrable chaotic 
systems, especially with a semiclassical limit, the expected exponential growth of the OTOC 
\begin{equation}
\label{eq:QuantumLE}
C_{\rho}(t)\sim e^{2 \lambda_Q t}
\end{equation}
has been observed and the quantum Lyapunov exponent $\lambda_Q$ has been found to be close to the
classical one. The exponential growth is observed till the log-time or the Ehrenfest time which 
scales as $\ln(1/h)/\lambda_{C}$ where $h$ is a scaled Planck's constant and $\lambda_C$ the classical
Lyapunov exponent.

The kicked top has been previously used in OTOC studies such as in \cite{Akshay2018,sieberer2019digital} and 
variations of it that break the permutation symmetry are beginning to be studied as well as potential models of 
``holography" \cite{yin2020quantum} as well as from the point of view of experimental realizations via NMR for example.
Previous studies of the kicked top OTOC were in the semiclassical limit of large $j$, wherein only numerical
results are accessible. It is of interest to ask how these properties manifest themselves in the solvable highly quantum
regime of small $j$ which are accessible to present day experiments. We are limited by short time scales and the exponential 
growths cannot be clearly observed in these cases. Yet it is intriguing to have exactly solvable cases wherein we may 
see such growth in a rudimentary form and study the transition to semiclassical regimes.
Due to our restriction to the permutation symmetric subspace, it is not 
possible to use a single qubit operator and we take the symmetric subspace projection of the
collective spin variable $A(0)= \sum_{i=1}^{2j}\sigma^z_i/2=J_z$ as the observable. 

\subsection{OTOC in 3 qubits: $j=3/2$}

For $j=3/2$, the 3 qubit case, this restriction takes the form of 
$J_z=(3/2)|000\kt \br 000|-(3/2) |111\kt \br 111| +(1/2) |W\kt \br W|-(1/2) |\overline{W}\kt \br \overline{W}|$.
Using the basis in Eq.~(\ref{eq:3qubit_basis_1},\ref{eq:3qubit_basis}) in  which the time evolution further block-diagonalizes and
noting that $J_z|\phi_1^{\pm}\kt = (3/2) |\phi_1^{\mp}\kt$, $J_z|\phi_2^{\pm}\kt = (1/2) |\phi_2^{\mp}\kt$,
we use
\begin{equation}
\label{eq:Jz}
J_z=\begin{pmatrix}
0_{2 \times 2 } &S\\S&0_{2 \times 2}
\end{pmatrix},
\;\; S=\frac{1}{2} \begin{pmatrix}
3 &0 \\0&1
\end{pmatrix}.
\end{equation}
This leads to 
\begin{equation}
J_z(n)=\mathcal{U}^{-n} J_z \mathcal{U}^n=\begin{pmatrix}
0 & \mathcal{U}_+^{-n} S \mathcal{U}_{-}^n\\
\mathcal{U}_-^{-n} S \mathcal{U}_{+}^n &0
\end{pmatrix}.
\end{equation}
Considering the case of the infinite temperature OTOC $C_{\infty}(n)$, we
separate it as
\begin{equation}
\label{eq:C2minusC4}
C_{\infty}(n)=C_2(n)-C_4(n),
\end{equation}
where $C_2(n)=\tr [J_z^2(n) J_z(0)^2]/4$ is the two-point correlator and $C_4(n)=\tr[ J_z(n) J_z(0) J_z(n)J_z(0)]/4$
is the four-point correlator which is out-of-time ordered. 
This leads to
\begin{subequations}
\label{eq:3qubitC2C4}
\begin{align}
C_2(n)&=\frac{1}{4}\left[ \tr(\mathcal{U}_+^{-n} S^2 \mathcal{U}_+^{n} S^2)+\tr(\mathcal{U}_-^{-n} S^2 \mathcal{U}_-^{n} S^2)\right]\\
C_4(n)&=\frac{1}{4}\left[ \tr(\mathcal{U}_+^{-n} S \mathcal{U}_-^{n} S)^2+\tr(\mathcal{U}_-^{-n} S \mathcal{U}_+^{n} S)^2\right].
\end{align}
\end{subequations}
Plugging in the elements of $\mathcal{U}_{\pm}^n$ from Eq.~(\ref{eq:Upluspowern}) and simplifications lead to 
\begin{equation}
\label{eq:Cinfty3qub}
\begin{split}
C_2(n)&=\frac{1}{16}\left( 41-32 |\beta_n|^2 \right)\\
C_4(n)&=(-1)^n \frac{1}{16} \left(41 -160 |\beta_n|^2+128 |\beta_n|^4 \right),
\end{split}
\end{equation}
where $\beta_n$ is given by Eq.~(\ref{eq:beta3qub}), and hence 
\begin{equation}
|\beta_n|^2=\frac{3}{4} U_{n-1}^2\left[ \frac{1}{2}\sin\left(\frac{\kappa_0}{3}\right) \right].
\end{equation} 
For small $\kappa_0$ when the dynamics is near-integrable these give
\begin{equation}
C_{\infty}(n)\approx \left\{ \begin{split}\frac{1}{6} n^2 \kappa_0^2 -\frac{13}{2592} n^4 \kappa_0^4 &\;\; n \; \text{even}\\
\frac{5}{8}+\frac{1}{288} (n^2-1)^2 \kappa_0^4 & \;\; n \; \text{odd} \end{split} \right.
\end{equation}
This shows a marked odd-even behaviour with the even time OTOC increasingly quadratically with time
at the lowest order. The odd-even effect is quite easily understood as for very small $\kappa_0$ the dynamics is essentially
one of rotation about the $y$ axis by $\pi/2$ and hence the $J_z$ operator with a concentration in the
$z$ direction is rotated practically to its negative at times $2 \,\text{mod}\, 4$ and to itself at times $0\, \text{mod}\, 4$ 
and hence almost commutes, but at times $1\, \text{mod}\, 4$ or $3 \, \text{mod}\, 4$ is 
concentrated on the $y$ and $-y$ directions and maximally fails to commute. Indeed the constant term $5/8$ is nothing but 
$-\tr[J_y,J_z]^2/4=\tr J_x^2/4$. A quadratic growth has also been observed in the Hadamard quantum walk \cite{SivaAL-2019} and we may expect 
a general power-law growth of the OTOC to be a general integrable and 
near-integrable feature \cite{Rozenbaum17,Prakash-2019} that we see in this small and solvable system exactly.

Now we turn attention to fixed and small times but for arbitrary values of the parameter $\kappa_0$. It follows from Eq.~(\ref{eq:C2minusC4}) and Eq.~(\ref{eq:Cinfty3qub}) that $C_{\infty}(1)=5/8$ irrespective of the value of $\kappa_0$, as $U_0(x)=1$. This shows no interesting dynamical
behaviour and the OTOC have a diffusive time scale over which the properties depend on the observable chosen as well. The next time steps 
already are of interest:
\begin{equation}
\begin{split}
C_{\infty}(2)&=6 \sin^2(\kappa_0/3)\left(1-\frac{3}{4} \sin^2(\kappa_0/3)\right)\\
C_{\infty}(3)&=\frac{5}{8}+18 \sin^4(\kappa_0/3)\left(1-\frac{1}{2}\sin^2(\kappa_0/3)\right)^2,
\end{split}
\end{equation}
$C_{\infty}(n)$ being a polynomial of order $4(n-1)$  in $\chi=\sin(\kappa_0/3)/2$.
\begin{figure}[h]
	\centering
	\includegraphics[width=0.8\linewidth]{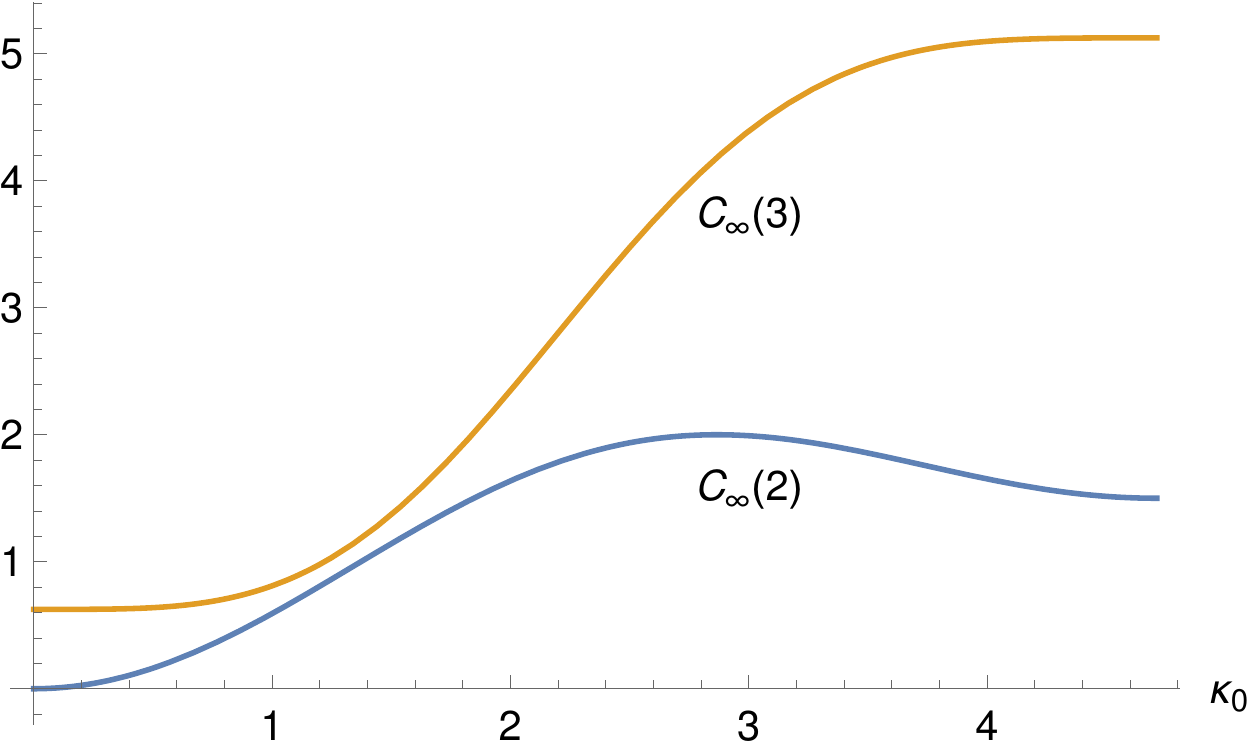}
	\caption{The OTOC for the 3 qubit kicked top at times $2$ and $3$ as a function of the chaos parameter $\kappa_0$. In all figures the observable used is $J_z$. Note the difference in the behavior 
	around $\kappa_0=0$, the near-integrable regime and also that the increase is monotonic at time $3$, and reaches a maximum at $\kappa_0=3 \pi/2$, when the top is essentially already fully chaotic.}
	\label{fig:OTOC2and3}
\end{figure}
The curves for $C_{\infty}(2)$ and $C_{\infty}(3)$ are shown in Fig.~(\ref{fig:OTOC2and3}) for convenience and 
we see that they increase with $\kappa_0$  and $C_{\infty}(3)$ is monotonically increasing over the entire range
of interest $\kappa_0 \in [0, 3 \pi/2]$, reaching the maximum value at $\kappa_0=3 \pi/2$.
\begin{figure}[h]
	\centering
	\includegraphics[width=0.8\linewidth]{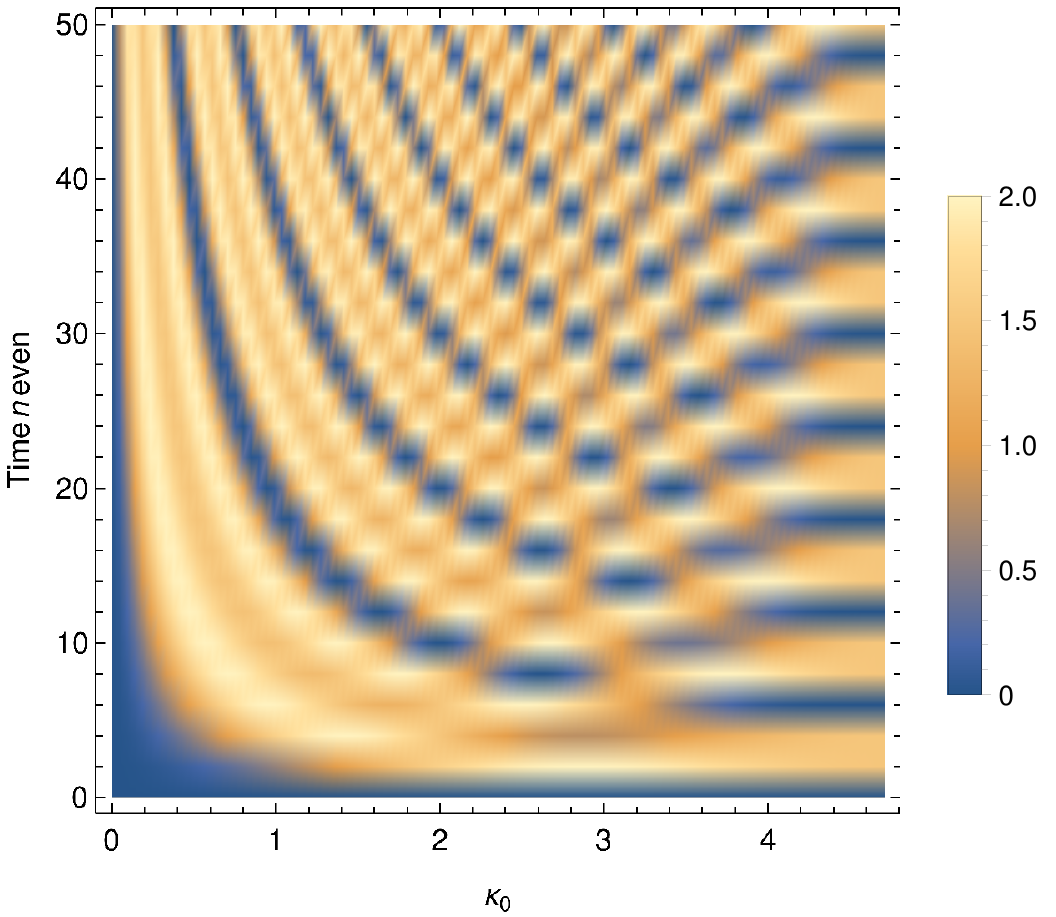}
	\includegraphics[width=0.8\linewidth]{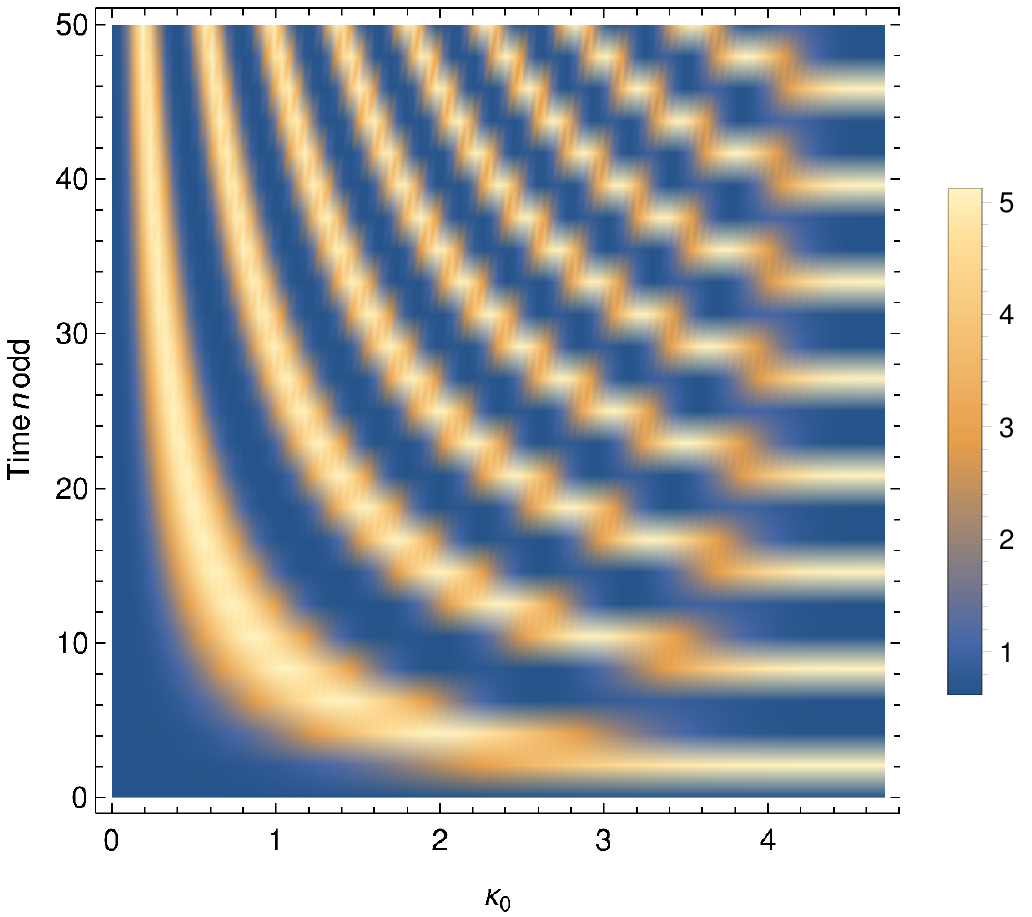}
	\caption{A density plot of the OTOC as function of time and the parameter $\kappa_0$ for 3 qubits, it is separated for 
	even and odd times for reasons explained in the text.}
	\label{fig:OTOC3evenodd}
\end{figure}
\begin{figure}[h]
	\centering
	\includegraphics[width=\linewidth]{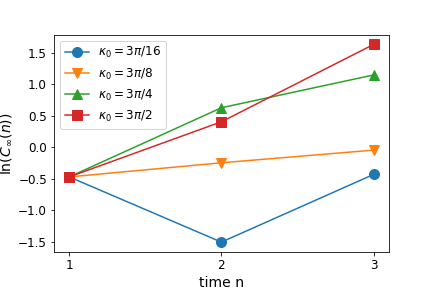}
	\caption{The OTOC is shown in linear-log scale for a few values of $\kappa_0$. The dynamics is predominantly chaotic at $\kappa_0=3 \pi/2$ and
	is reflected in what appears to be a near linear OTOC growth.}
	\label{fig:OTOCLogLin}
\end{figure}
A more global view is provided in Fig.~(\ref{fig:OTOC3evenodd}) where the OTOC for $j=3/2$ shown 
as a function of the time, split into even and odd ones, and the parameter $\kappa_0$. There would
be a periodicity beyond the value of $\kappa_0=3 \pi/2$, which provides an interesting boundary.
Exactly at this point, the classical dynamics is fairly chaotic and we do see a sharp increase in the OTOC
values for short times even in this small $j$ value. 

To give an indication of the growth, $\ln[C_{\infty}(n)]$ is 
plotted in Fig.~(\ref{fig:OTOCLogLin}) for $1\leq n \leq 3$. This has just three points, but the trend is clear and 
we may even interpret this as signs of the exponential growth of the OTOC that one expects in chaotic systems.
\begin{figure}[h]
	\centering
	\includegraphics[width=\linewidth]{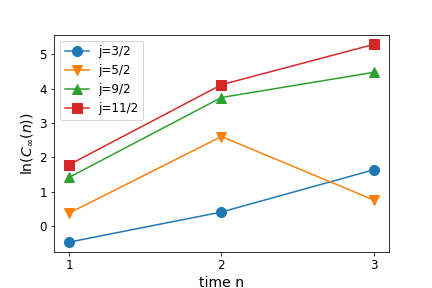}
	\caption{The OTOC in the linear-log scale, when $\kappa_0=3 \pi/2$ and the $j$ value is increased. The slope at $j=5/2$ already
	is well-saturated to those corresponding to larger $j$ values.}
	\label{fig:OTOC3qubkappa3piby2}
\end{figure}
To compare this with higher values of $j$, we show in Fig.~(\ref{fig:OTOC3qubkappa3piby2}) the case
for some larger values of the spin $j$, but with $\kappa_0=3 \pi/2$ in all the cases. We do see an increase and saturation 
in the slope with increasing $j$ values. It is interesting to observe from the same figure that with $j=5/2$ the OTOC slope has already 
saturated and hence at this value of the parameter, while $j=3/2$ is too low, $j=5/2$ may be just enough.
To explore this further we turn to the other solvable case of $j=2$ and compare it with
higher values of $j$, as well as study the peculiar case of $\kappa_0=\pi j$ for arbitrary $j$.

\subsection{OTOC in 4 qubits, $j=2$, and the peculiar case of $\kappa_0=\pi j$ for arbitrary $j$.}

The $4$ qubit case we reiterate can be qualitatively different from the case of $3$ as it has next-nearest neighbor interactions and is
a rudimentary non-integrable model. The calculations do not pose a serious problem as the unitary time evolution is still block-diagonalized
into utmost $2-$dimensional spaces, see Eq.~(\ref{eq:4qubitUpown}). the equations get a little bit more involved, but nevertheless can be
exactly solved, especially with the help of computer algebra. Skipping the details, we present the final results again separating the 
cases of different time parities. For time $n$ even we get
\begin{multline}
C_{\infty}(n)=\frac{1}{5}[ 34 -16\, |\beta_n|^2 \\-32 \,\text{Re} \left( \alpha_n^2 e^{in \kappa_0/4}\right)  
 -2 \cos(3 n \kappa_0/4) ],
\end{multline}
and for odd $n$, 
\begin{multline}
C_{\infty}(n)=\frac{1}{5}[ 25 -16\, |\beta_n|^2 \\-16 (-1)^{(n-1)/2} \, \text{Im}\left(\alpha_n e^{in \kappa_0/2}\right)].
\end{multline}
Here the $\alpha_n$ and $\beta_n$ involve the Chebyshev polynomials and are from Eq.~(\ref{eq:4QubitAlphaBeta}).
It follows that $C_{\infty}(1)=1$ irrespective of $\kappa_0$. 
Expressions for short times maybe explicitly extracted and for $n=2,3$ are
\begin{equation*}
\begin{split}
C_{\infty}(2)=&\frac{1}{5}\left(28-30\cos(\kappa_0/2)+6 \cos(\kappa_0)-4 \cos(3 \kappa_0/2)\right)\\
C_{\infty}(3)=&\frac{1}{10}\left(37-36 \cos(\kappa_0) +9 \cos(2 \kappa_0)\right).
\end{split}
\end{equation*}
While $C_{\infty}(2)$ is a monotonically increasing function for $0\leq \kappa_0 \leq 2 \pi$ and is a maximum at $\kappa_0=2\pi$,
$C_{\infty}(3)$ vanishes at this point having a maximum at $\kappa_0=\pi$. These special values of $\kappa_0$ correspond to $\pi j$ and $\pi j/2$. Notice that for $j=3/2$, $C(3)$ was a maximum at $\kappa_0=\pi j$ (see Fig.~(\ref{fig:OTOC2and3})), this difference between half-integer angular momenta and integer ones persists, and such features have also been noticed in entanglement before \cite{dogra19}. 

\begin{figure}[h]
	\centering
	\includegraphics[width=0.8\linewidth]{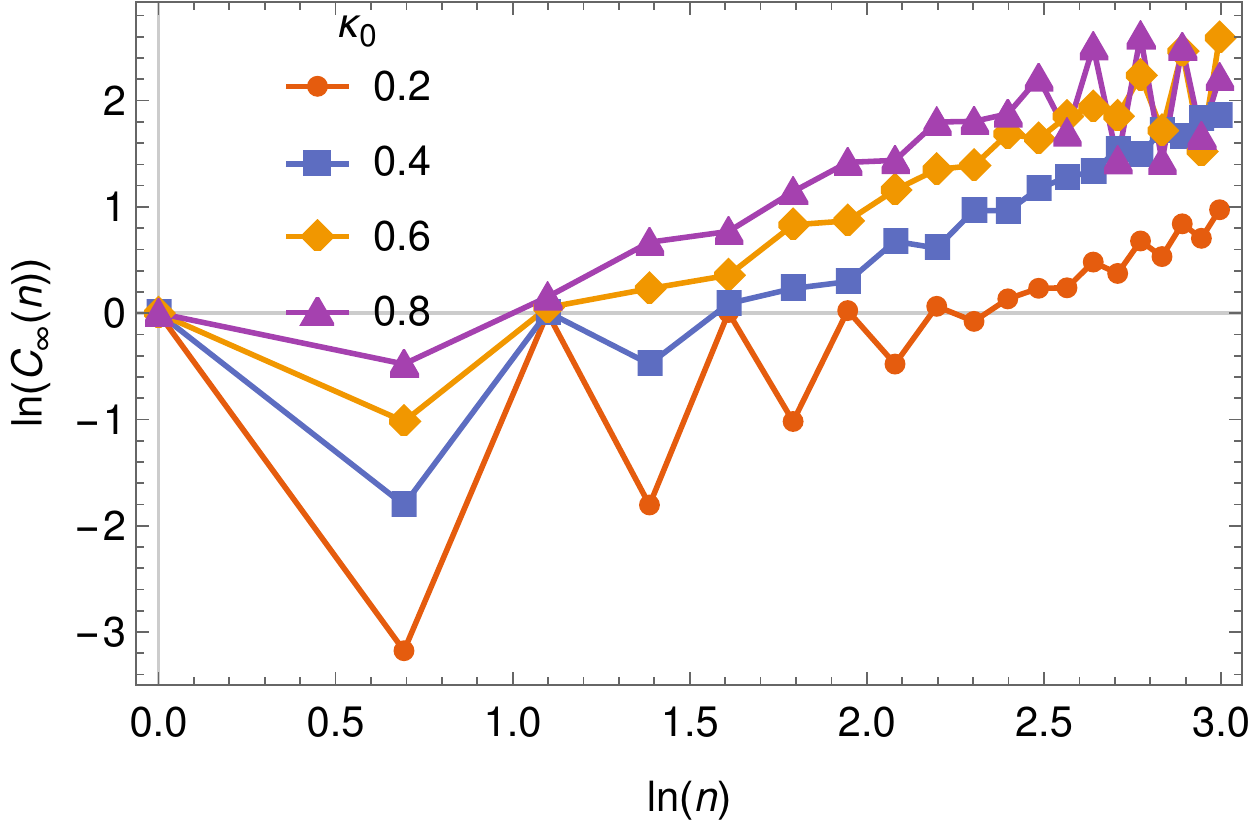}
	\caption{The 4 qubit OTOC growth in log-log scale for values of $\kappa_0$ when the dynamics is near-integrable. The growths
	are consistent with power-laws, taking into account the odd-even features in time.}
	\label{fig:OTOC4Qub_smallk}
\end{figure}
\begin{figure}[h]
	\centering
	\includegraphics[width=0.8\linewidth]{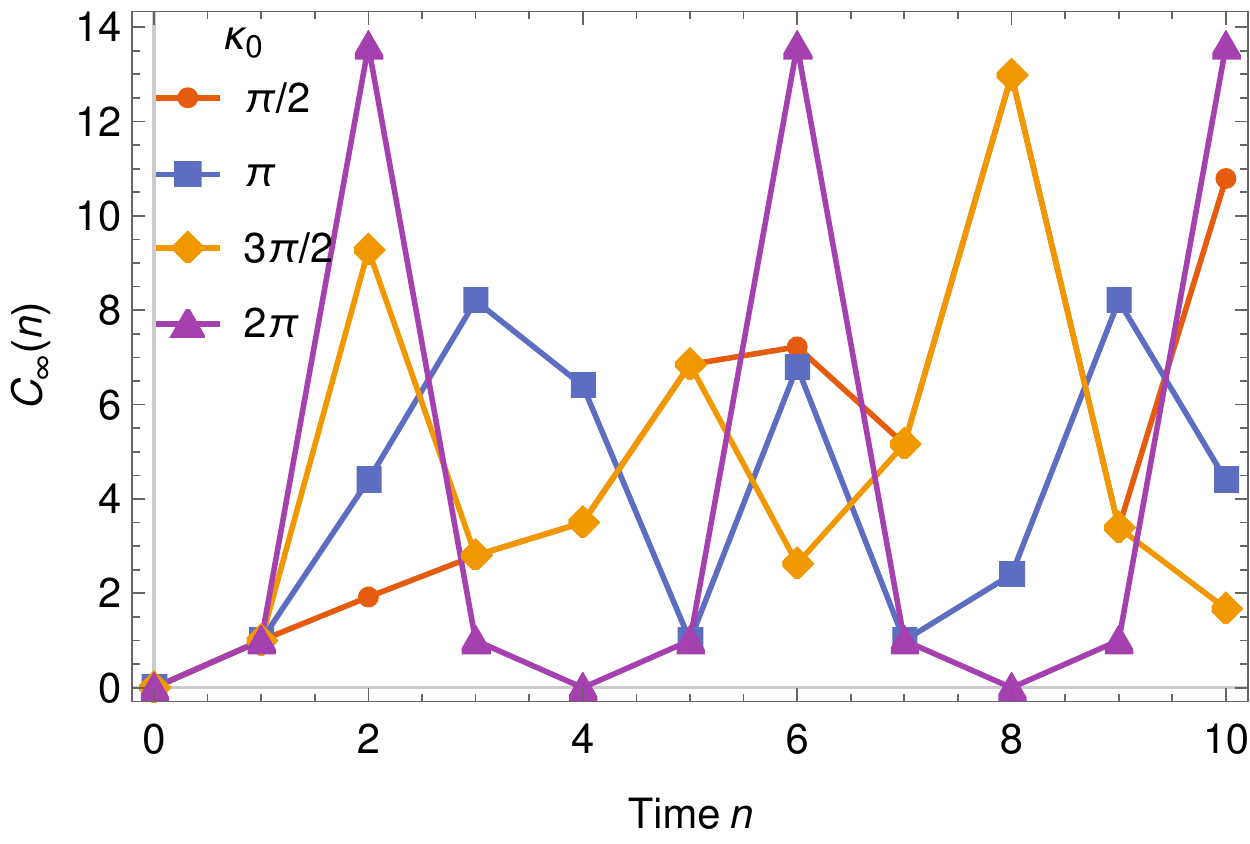}
	\caption{The 4 qubit OTOC for larger values of $\kappa_0$, the large growth at $\kappa_0=2 \pi$ is to be noted along with its periodicity.}
	\label{fig:OTOC4Qub_largek}
\end{figure}

For relatively small values of $\kappa_0$, when the classical system is near-integrable there is modest 
OTOC growth mostly governed by power laws as shown in Fig.~(\ref{fig:OTOC4Qub_smallk}). At large values of $\kappa_0$, the
OTOC grows rapidly, as seen in Fig.~(\ref{fig:OTOC4Qub_largek}) and then oscillates in an apparently irregular manner. 
\begin{figure}[h]
	\centering
	\includegraphics[width=0.8\linewidth]{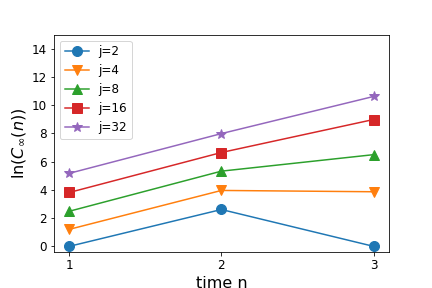}
	\caption{The $j=2$, 4 qubit OTOC at $\kappa_0=2 \pi$ compared with that of larger number of qubits, showing how the initial growth spurt is already reflecting the semiclassical Lyapunov exponent.}
	\label{fig:OTOC4Qub_pis}
\end{figure}
Of special interest again is $\kappa_0=2 \pi$, beyond which there is a symmetric behavior equivalent to a smaller value of $\kappa_0$ and hence certainly not reflecting any semi-classical property. For this case, it is amusing that the initial growth between $C_{\infty}(1)=1$ and $C_{\infty}(2)=68/5$, which is all that is there, in the sense that there is time-symmetry and periodicity beyond, already reflects the large $j$ growth of OTOC at $\kappa_0=2 \pi$.
The classical dynamics is highly chaotic at this parameter value and we may expect purely exponential growth of the OTOC. This is shown in Fig.~(\ref{fig:OTOC4Qub_pis}), where we only plot the first 3 time steps. Using the first 2 steps of the case $j=2$, we may be bold enough to 
find the quantum Lyapunov exponent of Eq.~(\ref{eq:QuantumLE}) as $0.5 \ln(68/5) \sim 1.3$ and compare with the classical value of $\lambda_{C}=\ln(\kappa_0)-1\sim 0.84$. We note of course that the classical exponent comes from an infinite time average and the kicked top, unlike the baker's or the cat map, is not a uniformly hyperbolic system. Thus it can hardly be expected that finite-time quantum properties from a particular observable reflect this
number exactly and we see that even for large $j$ the slope is not significantly changed towards the classical value. Thus it seems plausible that with only 4 qubits one can observe the exponential growth of the OTOC due to quantum chaos.

As the extreme case of $\kappa_0=\pi j$ registers the largest growth of the OTOC for the 3 and 4 qubit systems studied above, it is natural to
investigate this for an arbitrary value of $j$. In this case the Floquet unitary operator
\begin{equation}
\mathcal{U}=e^{-i \pi J_z^2/2}e^{-i \pi J_y/2}
\end{equation}
enjoys many special properties, that we intend to investigate in detail elsewhere. For integer $j$ values it is a {\it sum} of 4
pure rotations and in general, for integer $j$, we note that when $\kappa_0=\pi r/s$ where $r$ and $s$ are
relatively prime integers, 
\begin{align}
\mathcal{U}_{r,s}=&e^{-i r \pi J_z^2/2s}e^{-i \pi J_y/2}\\
&=\sum_{l=0}^{2s-1}a_l(r,s) e^{-i \pi l J_z/s}e^{-i \pi J_y/2}
\end{align}
where 
\begin{equation}
a_l(r,s)=\frac{1}{2s}\sum_{m=0}^{2s-1}e^{-i \pi m l/s}e^{-i \pi r m^2/s}
\end{equation}
are Gauss sums. A similar sum over $4s$ terms applies for half-integer $j$ values. We record them as possible
routes to implementing the kicked top experimentally when $\kappa_0$ is some rational multiple of $\pi$, as the
torsion is replaced by a sum of rotations. For the case of $j=2$, or $r=1$, $s=2$, we note that $\mathcal{U}^8=I$, where $I$ is identity. These maps remind one of the cat maps, whose quantum mechanics is exactly periodic.

For large value of $j$ we notice that the quantum-classical correspondence time, the Ehrenfest or log-time is $\sim \ln(2j+1)/\lambda_C=\ln(2j+1)/\ln(\pi j) \sim 1$. Thus we are at the true border of the correspondence and do not  expect to see classical effects for times beyond a few steps, however large $j$ may be, and indeed we find that only $n=1,2$ are unique and of interest. We find remarkably simple expressions for these:
\begin{equation}
\label{eq:PIJC}
\begin{split}
C_{\infty}(1)&=\frac{1}{6}j(j+1)\\
C_{\infty}(2)&=\frac{2}{15}j(j+1)(3 j^2+3j-1),
\end{split}
\end{equation}
they being related to squares and $4^{\text{th}}$ powers of integers. It reassuringly returns $1$ and $68/5$ for the case $j=2$ which we have discussed above.
This results in the quantum Lyapunov exponent of $\ln(C_{\infty}(2)/C_{\infty}(1))\sim \ln (j)+0.3$ which is to be compared with the classical one 
$\ln(\pi j)-1\sim \ln(j)+0.14$. Thus the principal growth of the two Lyapunov exponents are identical and we emphasize that this is in itself
quite a remarkable fact. Thus while this extreme case is highly special it does reflect the large classical chaos that underlies the system. Analysis for $\kappa_0$ other fractions of $\pi j$ are therefore of interest.

\section{Loschmidt echo and the kicked top}
\label{sec:Echo}

Loschmidt echo, as discussed above, is a quantifier of quantum chaos based on the overlap of a given state with itself when evolved by a perturbed and an exact Hamiltonian. In general, this depends on the choice of the initial state, 
 nature and magnitude of perturbation, degree of chaos.  
 To make the echo state independent, one can look at the decay by considering an average over initial states from Haar measure for finite dimensional systems, $\overline{F_d}(\kappa_0,\kappa_0',n)=\int d\ket{\psi_0} F_d(\kappa_0,\kappa_0',n,\ket{\psi_0})  $ and \cite{garcia2016lyapunov, zanardi2004purity} 
 \begin{equation}
\overline{F_d}(\kappa_0,\kappa_0',n)=\frac{1}{d(d+1)}(d+ \lvert\tr[\mathcal{U}^{-n}(\kappa_0) \mathcal{U}^{n}(\kappa_0')]\rvert^2)
 \end{equation}
 where $d$ is the dimension of the Hilbert space of the states. Essentially, the echo depends on the quantity $\lvert\tr[\mathcal{U}^{-n}(\kappa_0) \mathcal{U}^{n}(\kappa_0')]\rvert^2 $, which can be calculated easily to obtain, for the three qubit kicked top,
 \begin{equation}
 \overline{F_3}(\kappa_0,\kappa_0',n)=\frac{1}{5}(1+\lvert\alpha_n \tilde{\alpha^*_n}+\beta_n \tilde{\beta^*_n}+\beta^*_n \tilde{\beta_n}+\alpha^*_n \tilde{\alpha_n}\rvert^2)
 \end{equation}
 where $\tilde{\alpha_n}$ and $\tilde{\beta_n}$ are $\alpha_n(\kappa_0')$ and $\beta_n(\kappa_0')$ respectively and $\kappa_0'=\kappa_0+ \delta \kappa_0$. Here,  $\delta\kappa_0$ is the strength of perturbation.
 For the four qubit top, this gives 
 \begin{widetext}
 	\begin{equation}
 	\overline{F_4}(\kappa_0,\kappa_0',n)=\frac{1}{30}(5+\vert1+e^{\frac{in\delta \kappa_0}{4}}(\alpha_n \tilde{\alpha^*_n}+\beta_n \tilde{\beta^*_n}+\beta^*_n \tilde{\beta_n}+\alpha^*_n \tilde{\alpha_n} +2e^{\frac{3in\delta \kappa_0}{8}}(\cos^2(n\pi/2)+\sin^2 (n\pi/2)\cos(3\delta\kappa_0/8))\vert^2)
 	\end{equation}
 \end{widetext}
  Therefore, we have the exact expressions for the Loschmidt echo for the cases at hand and explore. 
 
 \begin{figure}[h]
 	\centering
 	\includegraphics[width=0.8\linewidth]{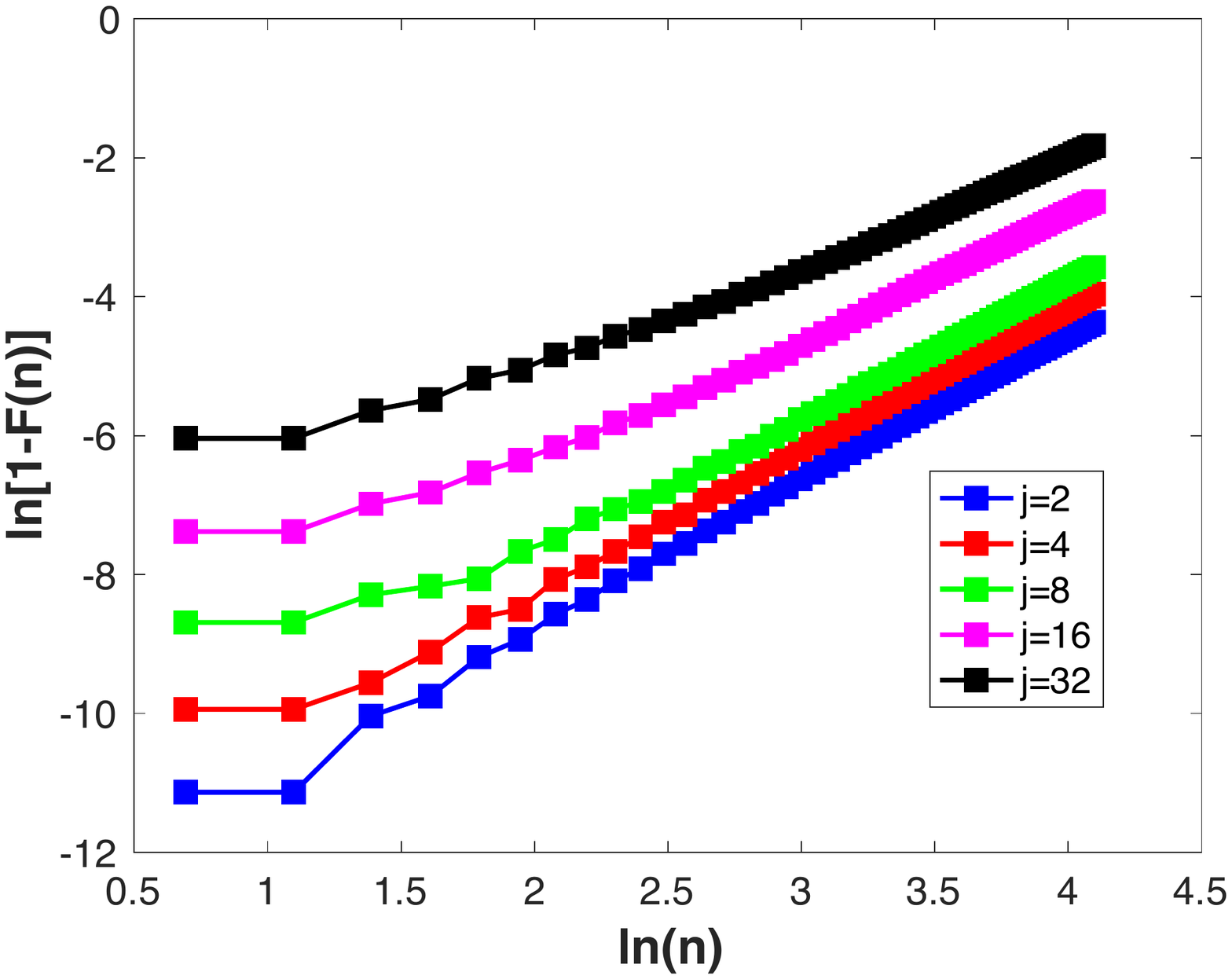}
 	\caption{The quadratic fall for the Loschmidt echo is shown with on a  log-log scale, when $\kappa_0= 2\pi$ for a few $j$ values including $j=2$. The perturbation strength is 0.01. }
 	\label{LE3}
 \end{figure}
 
 
  \begin{figure}[h]
 	\centering
 	\includegraphics[width=0.8\linewidth]{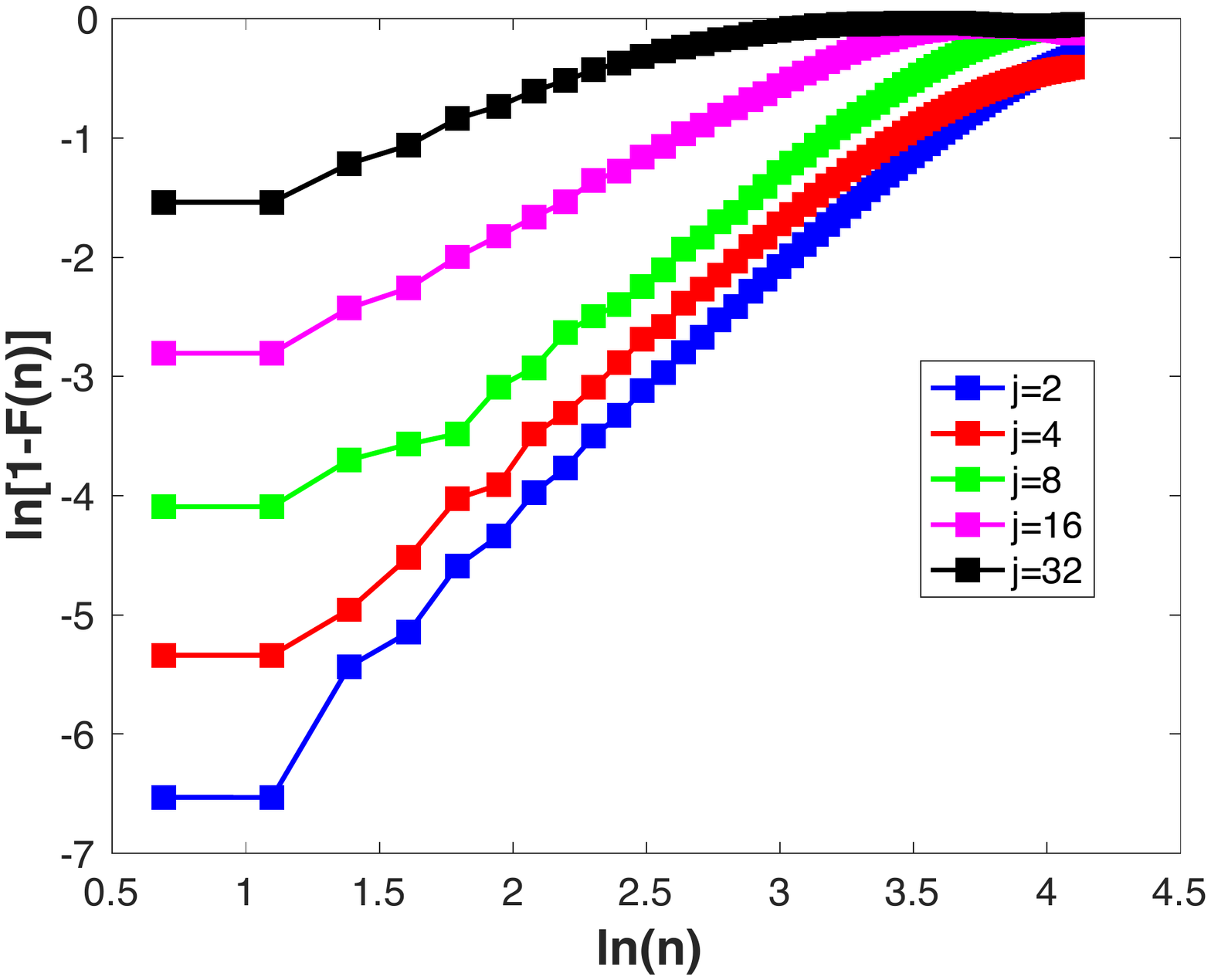}
 	\caption{The breakdown of quadratic fall for the Loschmidt echo is shown with on a  log-log scale, when $\kappa_0= 2\pi$ for a few $j$ values including $j=2$. The perturbation strength is 0.1.}
 	\label{LE4}
 \end{figure}
 
 

  \begin{figure}[h]
 	\centering
 	\includegraphics[width=0.8\linewidth]{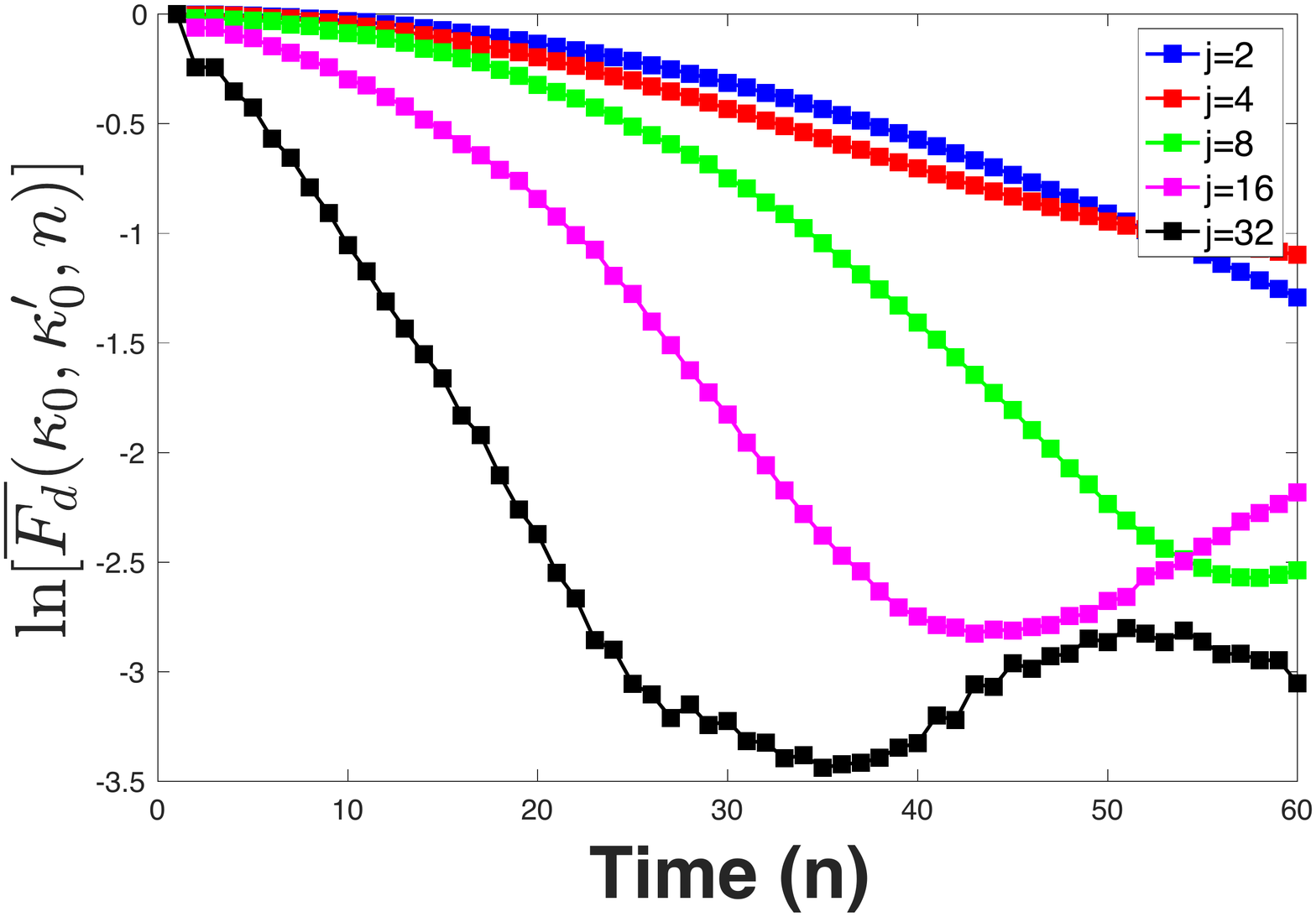}
 	\caption{Loschmidt decay on a linear log scale for some values of $j$. Perturbation strength, $\delta \kappa_0$, is 0.1 and $\kappa _0= 2\pi$. It can be seen that $j=32$ case is showing exponential decay - a forerunner of the Lyapunov decay. }
 			\label{LE5}
 \end{figure}

  \begin{figure}[h]
 	\centering
 	\includegraphics[width=0.8\linewidth]{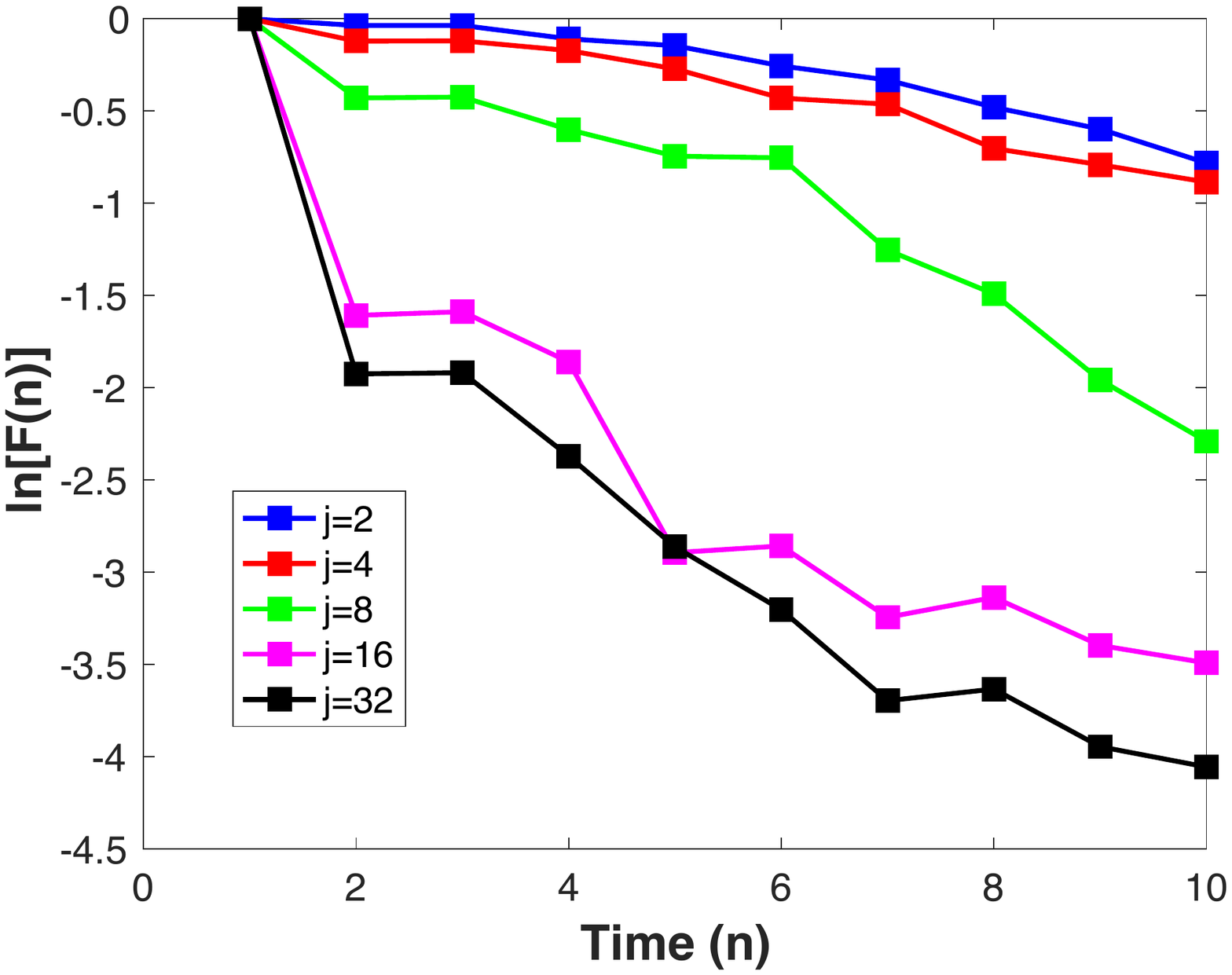}
 	\caption{Loschmidt decay on a linear log scale for some values of $j$. Perturbation strength, $\delta \kappa_0$, is $0.5$ and $\kappa_0 = 2\pi$. It can be seen that $j=16$ and  $j=32$ case is showing exponential decay - a forerunner of the Lyapunov decay. To extract the Lyapunov exponent, one needs to go for a much larger $j$.}
 	\label{LE6}
 \end{figure}
 
 \begin{figure}[h]
 	\centering
 	\includegraphics[width=0.8\linewidth]{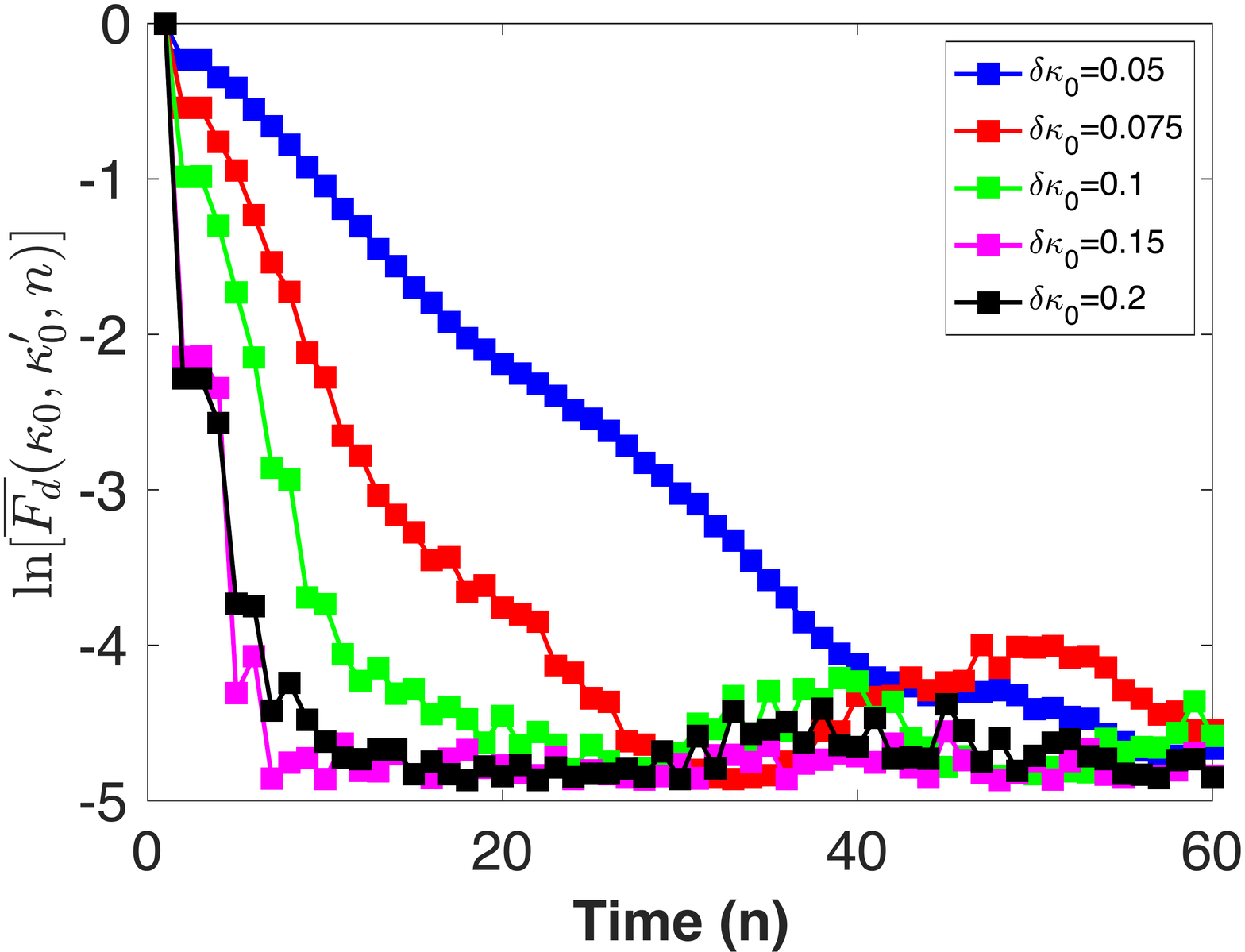}
 	\caption{Loschmidt decay on a linear log scale for some values of $\delta \kappa_0$ the perturbation strength.  $\kappa_0 = 2\pi$ and $j=64$. Increasing the perturbation strength results in saturation of the rate of exponential decay.}
 	\label{LE7}
 \end{figure}

 Figure (\ref{LE3}) shows that when the perturbation strength is low (order of $10^{-2}$), we see a Gaussian (quadratic in Log-Log plot) decay for the 3 and 4 qubit kicked top respectively. 
 Once the size of perturbation is increased, we see a departure from the quadratic decay as is shown in Fig (\ref{LE4}). 
 However, keeping the same perturbation strength, one observes an exponential decay in the echo as one increases the spin size for the kicked top as shown in (\ref{LE5}). Value of $j=32$ starts showing an exponential decay as evident on a log linear scale. As one increases the perturbation strength, as in Fig. (\ref{LE6}), one sees an exponential decay for $j=16$ and $j=32$ - a forerunner for the exponential decay
 that is the hallmark of Loschmidt decay in quantum chaotic systems. For large dimensional chaotic systems, as one increases the perturbation strength, there is a transisition from quadratic to exponential decay that saturates at the value given by the classical Lyapunov exponents \cite{garcia2016lyapunov}. We do see an antecedent of this decay in Fig. (\ref{LE7})  as on increasing perturbation strength, the decay rate saturates 
 to a fixed value.  Though we are still far from the semiclassical  quantum regime of large $j$, these numerical results serve as a precursor of Lyapunov decay for higher dimensional quantized chaotic Hamiltonians. 
 
 
\subsection{Fidelity decay for states}
\par
In this section, by considering the example of the 3 qubit kicked top system, we demonstrate how classical phase space features have an influence on the Loschmidt echo.  Analysis for four qubit states follows analogously.
Three-qubit permutation symmetric 
initial states used are coherent states located at 
\begin{eqnarray}
X_{0}&=&\sin\theta_0 \cos\phi_0, \nonumber \\ 
Y_{0}&=&\sin \theta_0 \sin\phi_0, \nonumber \\ 
Z_{0}&=&\cos \theta_0,
\end{eqnarray}
on the phase space sphere and given by~\cite{Glauber,Puri},
\begin{equation}
\ket{\psi_0}=|\theta_0,\phi_0\kt = \otimes^{2j} 
(\cos(\theta_0/2) |0\kt + e^{-i \phi_0} \sin(\theta_0/2) |1\kt).
\end{equation}

We  study time evolution and fidelity decay of two completely different three-qubit states
 ((i) $|0,0\kt$ and (ii) $|\pi/2, -\pi/2\kt$), shown in Fig. (\ref{fig:classical}).
The
coherent state at $|0,0\kt$ for three qubits is  $\otimes^3|0\kt.$ It is on a period-4 orbit in the classical phase space and is represented  with a red square
in~\ref{fig:classical}.  $\otimes^3|+\kt_y$ corresponds 
to the coherent state at $|\pi/2,-\pi/2\kt$,
which is a fixed point as per regular classical phase space 
structure, and eventually becomes unstable as we move 
from regular to mixed phase space, shown by a red circle in~Fig.(\ref{fig:classical}).
Let us consider the state on the period-4 orbit, corresponding to the 
coherent state at $|0,0\kt$ which is $\otimes^3|0\kt$.
\begin{equation}
\label{eq29}
\begin{split}
&\mathcal{U}^{n} |000\rangle \equiv |\psi_n \kt =\frac{1}{2}e^{-i n \left(\frac{3 \pi}{4}+\kappa\right)}\left\lbrace (1+i^n) \left(
\alpha_n |000\rangle  \right.  \right. \\ & \left. \left. + i \beta_n |\overline{W} \rangle
\right) + (1-i^n) \left( i \alpha_n |111\rangle - \beta_n |W \rangle
\right) \right\rbrace. 
\end{split} 
\end{equation}

%
Loschmidt decay can be computed by looking at the overlap of this state with another, evolved with a unitary of slightly different chaoticity parameter $\kappa_0'.$ 
\begin{align*}
F_3(\kappa_0,\kappa_0',n,\ket{\psi_0})&=|\bra{000} \mathcal{U}^{-n} (\kappa_0)\mathcal{U}^{n}(\kappa_0')  \ket{000}|^2\\
&= |\alpha_n^* \tilde{\alpha_n}+\beta_n^*\tilde{\beta_n} |^2 \numberthis
\end{align*}

Expansion in powers of $\delta \kappa_0$ at $\kappa_0=3\pi/2$ yields the quadratic term as the leading term that is non-zero for 
$n=4$ and beyond. This explains the extremely slow fall in fidelity for this state at $\kappa_0 = 3\pi/2$.  In contrast, the quadratic term in the expansion of  decay for $\kappa_0 = 0$  becomes non-zero starting with $n=1$.

\begin{figure}[h!]
	\centering
	\includegraphics[width=0.8\linewidth]{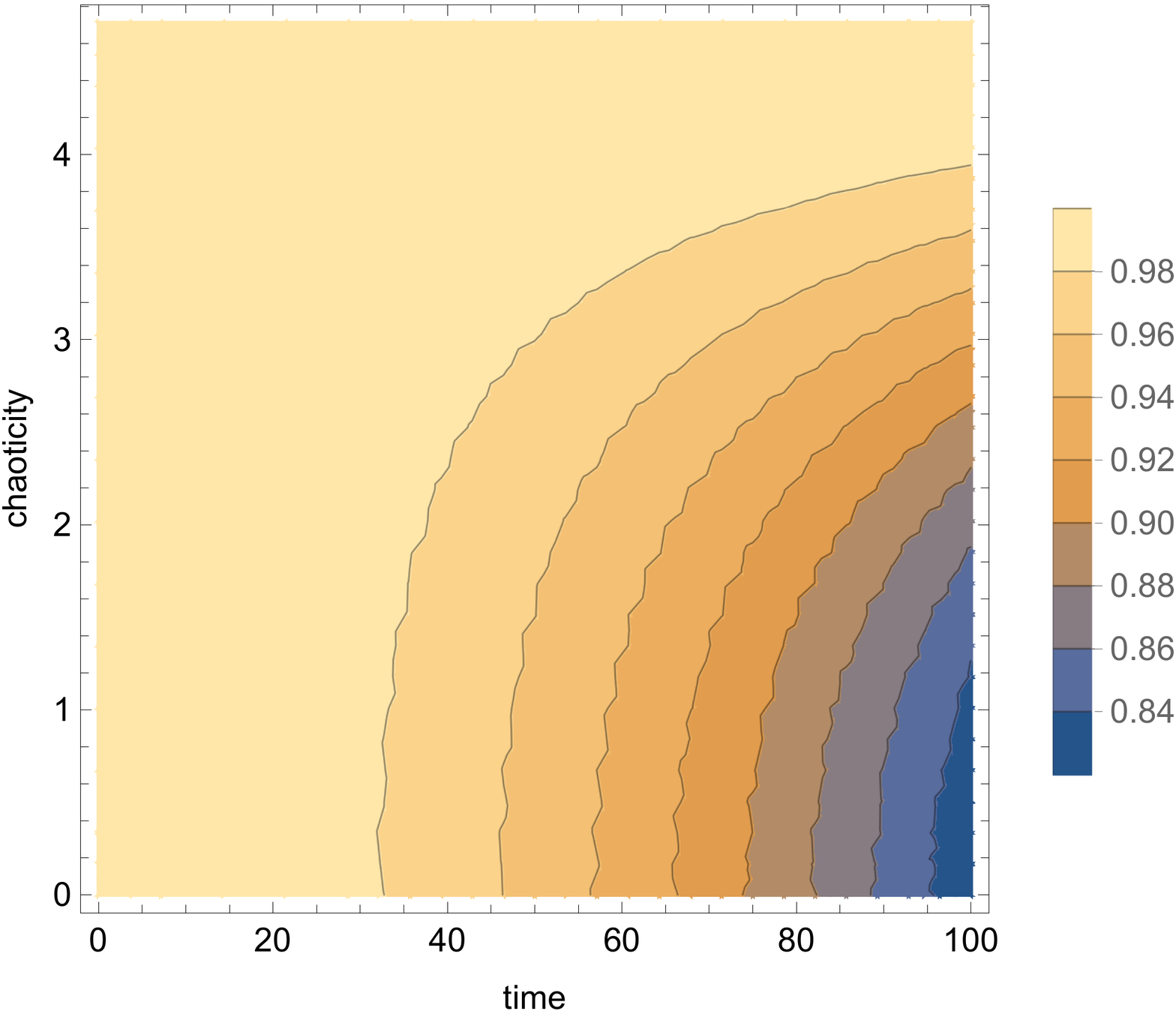}
	\caption{Loschmidt decay for $\ket{000}$ with respect to the chaoticity parameter $\kappa_0 \in [0, 3 \pi/2]$ and time $n$.  Perturbation strength is fixed at $0.005.$}
	\label{LE000}
\end{figure}

Figure (\ref{LE000}) interestingly shows somewhat counter-intuitive behavior of the decay of Loschmidt echo 
with chaos. At first, it appears, more chaos leads to less echo decay for a coherent wave packet starting at 
$|0,0\kt$. However, the state $|0,0\kt$ is on a period 4 orbit and will rapidly become delocalized with support over the period 4 phase space points. Fidelity decay for delocalized states having a high participation ratio is in general inversely correlated with the degree of chaos \cite{gorin2006dynamics}. 
As a contrast, consider the three-qubit state, $|\psi_0\rangle=|+++\rangle$, corresponding to a fixed point of the map, where
$|+\rangle=\frac{1}{\sqrt{2}}(|0\rangle+i|1\rangle)$
is an eigenvector of $\sigma_y$ with eigenvalue $+1$. This state delocalizes when the fixed point loses stabiity and the echo decay increases with the increase of chaos ($\kappa_0 \in [0, 3 \pi/2]$) in the system. 
When the initial state is $\otimes^3|+\kt_y=(|\phi_1^+\kt +\sqrt{3} i |\phi_2^+\kt)/2$, 
corresponding to the coherent state at $|\pi/2,-\pi/2\kt$, the evolution lies entirely in the positive 
parity sector. We have, $\mathcal{U}^n|+++\kt_y$ equal to
\begin{align*}
|\psi_n\kt 
=\frac{1}{2} e^{-i n \left(\frac{\pi}{4}+\kappa\right)}\big[ (\alpha_n-i\sqrt{3}\beta_n^{*}) |\phi_1^{+} \rangle+(\beta_n+i\sqrt{3}\alpha_n^{*})|\phi_2^{+} \rangle \big].
\label{eq:fixedptstate}
\end{align*}
Defining $\gamma_n=(\alpha_n-i\sqrt{3}\beta_n^{*})/2$ and $\delta_n= (\beta_n+i\sqrt{3}\alpha_n^{*})/2$,
we can obtain the fidelity decay expression at time $n$ as before.
\begin{align*}
F(\kappa_0,\kappa_0',n, \ket{\psi_0})&=|\bra{+++} \mathcal{U}^{-n} (\kappa_0)\mathcal{U}^{n}(\kappa_0')  \ket{+++}|^2\\
&= |\gamma_n^*\tilde{\gamma_n} +\delta_n^*\tilde{\delta_n} |^2 \numberthis
\end{align*}
where $\tilde{\gamma_n}$ and  $\tilde{\delta_n}$ are $\gamma_n(\kappa'_0)$ and $\delta_n(\kappa_0')$ respectively.

\begin{figure}[h]
	\centering
	\includegraphics[width=0.8\linewidth]{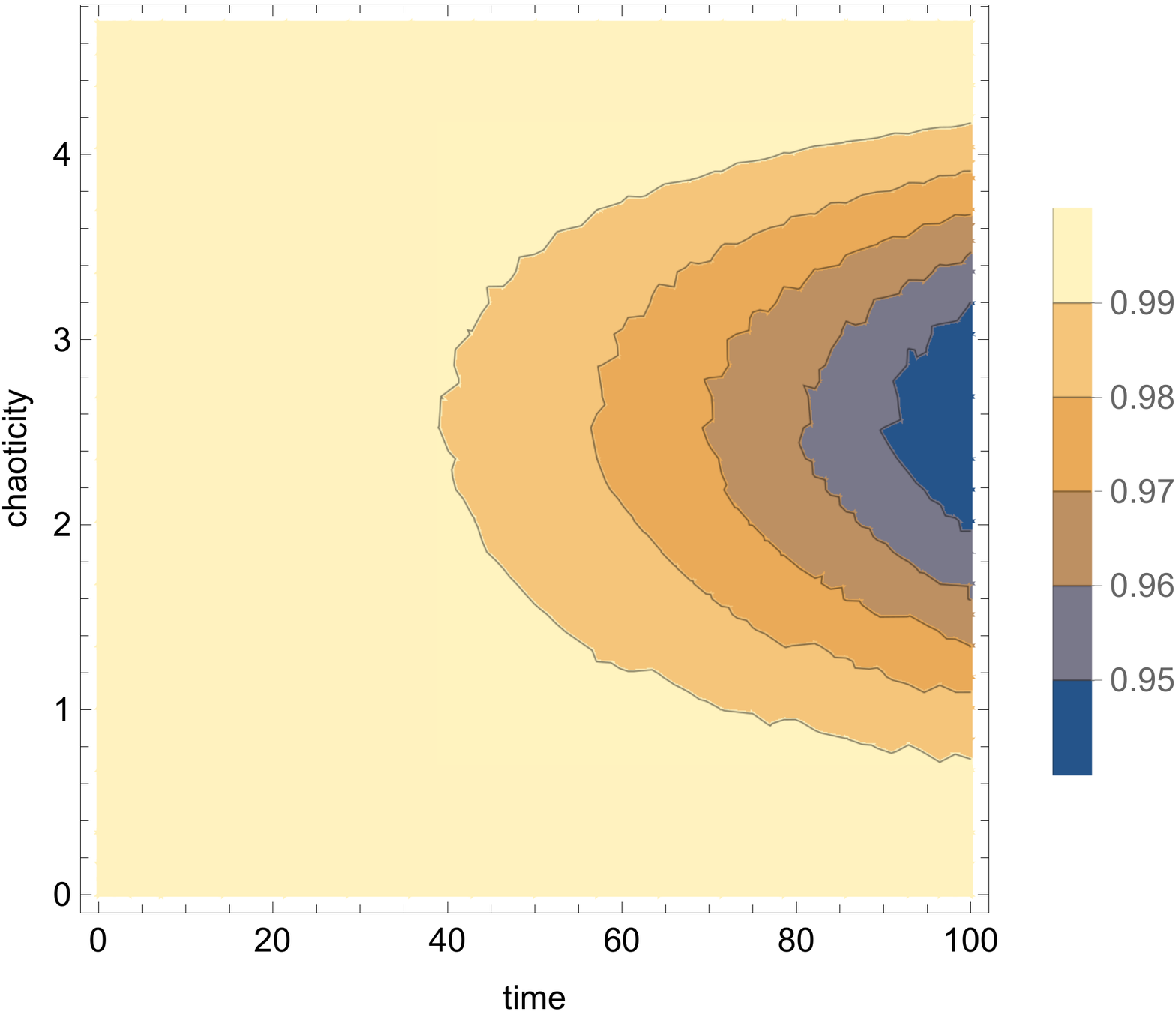}
	\caption{Loschmidt decay for $\ket{+++}$ with respect to the chaoticity parameter $\kappa_0 \in [0, 3 \pi/2]$ and time $n$.  Perturbation strength is fixed at $0.005.$}
\end{figure}
Decay of Loschmidt echo for the $\ket{+++}$ for small perturbations follows the quadratic decay and also
increases with the increase of chaos ($\kappa_0 \in [0, 3 \pi/2]$) in the system.

\section{Summary and Discussion}
\label{sec:Conc}

Quantum chaos investigates the footprints of classical chaos in the quantum world. We posed an intriguing question - how deep in the quantum regime one can hope to find these signatures? In our work, we addressed this question with a provocative answer - we find signatures of classical Lyapunov exponents
as captured by OTOCs even in quantum systems consisting of as few as 3 and 4 qubits. Our results for Loschmidt echo, another quantifier of chaos based on sensitive dependence of a system to perturbations in dynamics, 
suggest a more feeble signature of chaos for the kicked top with lower values of angular momentum. Through numerical study, we have shown that one needs to go to 
sufficiently high quantum numbers to see a forerunner to the exponential Lyapunov decay in the Loschmidt echo. However, for certain initial states, we do see the effects of delocalization, periodic orbits, and chaos in the decay of the echo signal in deep quantum regime of 3 and 4 qubit kicked top. How do these states fare under environmental decoherence would be an interesting future direction to explore.

Recent studies involving a related concept, the Adiabatic Guage Potential (AGP) which is the generator of adiabatic deformations between eigenstates, serves as a probe to detect chaos in systems with large Hilbert spaces \cite{PhysRevX.10.041017}. An interesting direction for the future is to compare the effectiveness of AGP with that of Loschmidt echo in detecting chaos. 

One  interesting observation from our work was the case of $\kappa_0=\pi j$, the chaoticity parameter for the kicked top. As we saw, for the value of $j=2$, the Floequet operator in this case, has interesting decomposition in terms of 
 {\it sum} of 4
pure rotations and similar sum exists for $\kappa_0=\pi r/s$ with $r$ and $s$ relatively prime to each other. 
On the one hand, this paves way for some experiments where the nonlinear twist is replaced by a sum of rotations. On the other hand, this gives us some insights into the origin of chaos and complexity in a system
with a classical limit of just two degrees of freedom.
It is also worth noting that it is very rare that systems exhibiting signatures of chaos are exactly solvable.
A conservative system with as many constants of motion as its degrees of freedom is said to be integrable and its dynamics regular. In the quantum world, these constants of motion become operators that commute with the Hamiltonian. Lack of sufficient constants of motion leads to non-integrability and the random matrix conjecture in the quantum domain. Exactly solvable systems give us a reference to study departure from integrability and transition to chaos upon the introduction of perturbations breaking the necessary symmetries via the KAM theorem. Our work paves way for the search for more systems that are ``chaotic" yet solvable. For example, a system of coupled kicked tops \cite{tmd08}, which consists of two spins coupled via hyperfine interactions and one of them periodically kicked can be made to have connections with a many-body model considering a large spin as a collection of spin 1/2 particles. We hope our work will be interesting and useful to the quantum chaos community as well as experimentalists.


\begin{thebibliography}{86}
\expandafter\ifx\csname natexlab\endcsname\relax\def\natexlab#1{#1}\fi
\expandafter\ifx\csname bibnamefont\endcsname\relax
  \def\bibnamefont#1{#1}\fi
\expandafter\ifx\csname bibfnamefont\endcsname\relax
  \def\bibfnamefont#1{#1}\fi
\expandafter\ifx\csname citenamefont\endcsname\relax
  \def\citenamefont#1{#1}\fi
\expandafter\ifx\csname url\endcsname\relax
  \def\url#1{\texttt{#1}}\fi
\expandafter\ifx\csname urlprefix\endcsname\relax\def\urlprefix{URL }\fi
\providecommand{\bibinfo}[2]{#2}
\providecommand{\eprint}[2][]{\url{#2}}

\bibitem[{\citenamefont{Neill et~al.}(2016)\citenamefont{Neill, Roushan, Fang,
  Chen, Kolodrubetz, Chen, Megrant, Barends, Campbell, Chiaro
  et~al.}}]{Neill16}
\bibinfo{author}{\bibfnamefont{C.}~\bibnamefont{Neill}},
  \bibinfo{author}{\bibfnamefont{P.}~\bibnamefont{Roushan}},
  \bibinfo{author}{\bibfnamefont{M.}~\bibnamefont{Fang}},
  \bibinfo{author}{\bibfnamefont{Y.}~\bibnamefont{Chen}},
  \bibinfo{author}{\bibfnamefont{M.}~\bibnamefont{Kolodrubetz}},
  \bibinfo{author}{\bibfnamefont{Z.}~\bibnamefont{Chen}},
  \bibinfo{author}{\bibfnamefont{A.}~\bibnamefont{Megrant}},
  \bibinfo{author}{\bibfnamefont{R.}~\bibnamefont{Barends}},
  \bibinfo{author}{\bibfnamefont{B.}~\bibnamefont{Campbell}},
  \bibinfo{author}{\bibfnamefont{B.}~\bibnamefont{Chiaro}},
  \bibnamefont{et~al.}, \bibinfo{journal}{Nature Physics}
  \textbf{\bibinfo{volume}{12}}, \bibinfo{pages}{1037} (\bibinfo{year}{2016}).

\bibitem[{\citenamefont{Kaufman et~al.}(2016)\citenamefont{Kaufman, Tai, Lukin,
  Rispoli, Schittko, Preiss, and Greiner}}]{Kaufman2016}
\bibinfo{author}{\bibfnamefont{A.~M.} \bibnamefont{Kaufman}},
  \bibinfo{author}{\bibfnamefont{M.~E.} \bibnamefont{Tai}},
  \bibinfo{author}{\bibfnamefont{A.}~\bibnamefont{Lukin}},
  \bibinfo{author}{\bibfnamefont{M.}~\bibnamefont{Rispoli}},
  \bibinfo{author}{\bibfnamefont{R.}~\bibnamefont{Schittko}},
  \bibinfo{author}{\bibfnamefont{P.~M.} \bibnamefont{Preiss}},
  \bibnamefont{and} \bibinfo{author}{\bibfnamefont{M.}~\bibnamefont{Greiner}},
  \bibinfo{journal}{Science} \textbf{\bibinfo{volume}{353}},
  \bibinfo{pages}{794} (\bibinfo{year}{2016}), ISSN \bibinfo{issn}{0036-8075},
  \eprint{http://science.sciencemag.org/content/353/6301/794.full.pdf},
  \urlprefix\url{http://science.sciencemag.org/content/353/6301/794}.

\bibitem[{\citenamefont{Miller and Sarkar}(1999)}]{MillerSarkar}
\bibinfo{author}{\bibfnamefont{P.~A.} \bibnamefont{Miller}} \bibnamefont{and}
  \bibinfo{author}{\bibfnamefont{S.}~\bibnamefont{Sarkar}},
  \bibinfo{journal}{Phys. Rev. E} \textbf{\bibinfo{volume}{60}},
  \bibinfo{pages}{1542} (\bibinfo{year}{1999}).

\bibitem[{\citenamefont{Lakshminarayan}(2001)}]{Lakshminarayan}
\bibinfo{author}{\bibfnamefont{A.}~\bibnamefont{Lakshminarayan}},
  \bibinfo{journal}{Phys. Rev. E} \textbf{\bibinfo{volume}{64}},
  \bibinfo{pages}{036207} (\bibinfo{year}{2001}).

\bibitem[{\citenamefont{Bandyopadhyay and
  Lakshminarayan}(2002)}]{BandyopadhyayArul2002}
\bibinfo{author}{\bibfnamefont{J.~N.} \bibnamefont{Bandyopadhyay}}
  \bibnamefont{and}
  \bibinfo{author}{\bibfnamefont{A.}~\bibnamefont{Lakshminarayan}},
  \bibinfo{journal}{Phys. Rev. Lett.} \textbf{\bibinfo{volume}{89}},
  \bibinfo{pages}{060402} (\bibinfo{year}{2002}).

\bibitem[{\citenamefont{Tanaka et~al.}(2002)\citenamefont{Tanaka, Fujisaki, and
  Miyadera}}]{Tanaka-2002}
\bibinfo{author}{\bibfnamefont{A.}~\bibnamefont{Tanaka}},
  \bibinfo{author}{\bibfnamefont{H.}~\bibnamefont{Fujisaki}}, \bibnamefont{and}
  \bibinfo{author}{\bibfnamefont{T.}~\bibnamefont{Miyadera}},
  \bibinfo{journal}{Phys. Rev. E} \textbf{\bibinfo{volume}{66}},
  \bibinfo{pages}{045201} (\bibinfo{year}{2002}),
  \urlprefix\url{https://link.aps.org/doi/10.1103/PhysRevE.66.045201}.

\bibitem[{\citenamefont{Lakshminarayan and Subrahmanyam}(2003)}]{LakSub2003}
\bibinfo{author}{\bibfnamefont{A.}~\bibnamefont{Lakshminarayan}}
  \bibnamefont{and}
  \bibinfo{author}{\bibfnamefont{V.}~\bibnamefont{Subrahmanyam}},
  \bibinfo{journal}{Phys. Rev. A} \textbf{\bibinfo{volume}{67}},
  \bibinfo{pages}{052304} (\bibinfo{year}{2003}).

\bibitem[{\citenamefont{Bandyopadhyay and
  Lakshminarayan}(2004)}]{Bandyopadhyay04}
\bibinfo{author}{\bibfnamefont{J.~N.} \bibnamefont{Bandyopadhyay}}
  \bibnamefont{and}
  \bibinfo{author}{\bibfnamefont{A.}~\bibnamefont{Lakshminarayan}},
  \bibinfo{journal}{Phys. Rev. E} \textbf{\bibinfo{volume}{69}},
  \bibinfo{pages}{016201} (\bibinfo{year}{2004}).

\bibitem[{\citenamefont{Ghose and Sanders}(2004)}]{Ghose2004}
\bibinfo{author}{\bibfnamefont{S.}~\bibnamefont{Ghose}} \bibnamefont{and}
  \bibinfo{author}{\bibfnamefont{B.~C.} \bibnamefont{Sanders}},
  \bibinfo{journal}{Phys. Rev. A} \textbf{\bibinfo{volume}{70}},
  \bibinfo{pages}{062315} (\bibinfo{year}{2004}).

\bibitem[{\citenamefont{Lakshminarayan and Subrahmanyam}(2005)}]{ArulSub2005}
\bibinfo{author}{\bibfnamefont{A.}~\bibnamefont{Lakshminarayan}}
  \bibnamefont{and}
  \bibinfo{author}{\bibfnamefont{V.}~\bibnamefont{Subrahmanyam}},
  \bibinfo{journal}{Phys. Rev. A} \textbf{\bibinfo{volume}{71}},
  \bibinfo{pages}{062334} (\bibinfo{year}{2005}),
  \urlprefix\url{https://link.aps.org/doi/10.1103/PhysRevA.71.062334}.

\bibitem[{\citenamefont{Trail et~al.}(2008{\natexlab{a}})\citenamefont{Trail,
  Madhok, and Deutsch}}]{trail2008entanglement}
\bibinfo{author}{\bibfnamefont{C.~M.} \bibnamefont{Trail}},
  \bibinfo{author}{\bibfnamefont{V.}~\bibnamefont{Madhok}}, \bibnamefont{and}
  \bibinfo{author}{\bibfnamefont{I.~H.} \bibnamefont{Deutsch}},
  \bibinfo{journal}{Physical Review E} \textbf{\bibinfo{volume}{78}},
  \bibinfo{pages}{046211} (\bibinfo{year}{2008}{\natexlab{a}}).

\bibitem[{\citenamefont{Lombardi and Matzkin}(2011)}]{LombardiMatzkin2011}
\bibinfo{author}{\bibfnamefont{M.}~\bibnamefont{Lombardi}} \bibnamefont{and}
  \bibinfo{author}{\bibfnamefont{A.}~\bibnamefont{Matzkin}},
  \bibinfo{journal}{Phys. Rev. E} \textbf{\bibinfo{volume}{83}},
  \bibinfo{pages}{016207} (\bibinfo{year}{2011}),
  \urlprefix\url{https://link.aps.org/doi/10.1103/PhysRevE.83.016207}.

\bibitem[{\citenamefont{Madhok et~al.}(2014)\citenamefont{Madhok, Riofr\'{\i}o,
  Ghose, and Deutsch}}]{mrgi14}
\bibinfo{author}{\bibfnamefont{V.}~\bibnamefont{Madhok}},
  \bibinfo{author}{\bibfnamefont{C.~A.} \bibnamefont{Riofr\'{\i}o}},
  \bibinfo{author}{\bibfnamefont{S.}~\bibnamefont{Ghose}}, \bibnamefont{and}
  \bibinfo{author}{\bibfnamefont{I.~H.} \bibnamefont{Deutsch}},
  \bibinfo{journal}{Phys. Rev. Lett.} \textbf{\bibinfo{volume}{112}},
  \bibinfo{pages}{014102} (\bibinfo{year}{2014}).

\bibitem[{\citenamefont{Madhok et~al.}(2015)\citenamefont{Madhok, Gupta,
  Trottier, and Ghose}}]{madhok2015signatures}
\bibinfo{author}{\bibfnamefont{V.}~\bibnamefont{Madhok}},
  \bibinfo{author}{\bibfnamefont{V.}~\bibnamefont{Gupta}},
  \bibinfo{author}{\bibfnamefont{D.-A.} \bibnamefont{Trottier}},
  \bibnamefont{and} \bibinfo{author}{\bibfnamefont{S.}~\bibnamefont{Ghose}},
  \bibinfo{journal}{Physical Review E} \textbf{\bibinfo{volume}{91}},
  \bibinfo{pages}{032906} (\bibinfo{year}{2015}).

\bibitem[{\citenamefont{Madhok et~al.}(2018)\citenamefont{Madhok, Dogra, and
  Lakshminarayan}}]{Madhok2018_corr}
\bibinfo{author}{\bibfnamefont{V.}~\bibnamefont{Madhok}},
  \bibinfo{author}{\bibfnamefont{S.}~\bibnamefont{Dogra}}, \bibnamefont{and}
  \bibinfo{author}{\bibfnamefont{A.}~\bibnamefont{Lakshminarayan}},
  \bibinfo{journal}{Optics Communications} \textbf{\bibinfo{volume}{420}},
  \bibinfo{pages}{189 } (\bibinfo{year}{2018}), ISSN \bibinfo{issn}{0030-4018},
  \urlprefix\url{https://www.sciencedirect.com/science/article/pii/S0030401818302542}.

\bibitem[{\citenamefont{Piga et~al.}(2019)\citenamefont{Piga, Lewenstein, and
  Quach}}]{Maciej-2019}
\bibinfo{author}{\bibfnamefont{A.}~\bibnamefont{Piga}},
  \bibinfo{author}{\bibfnamefont{M.}~\bibnamefont{Lewenstein}},
  \bibnamefont{and} \bibinfo{author}{\bibfnamefont{J.~Q.} \bibnamefont{Quach}},
  \bibinfo{journal}{Phys. Rev. E} \textbf{\bibinfo{volume}{99}},
  \bibinfo{pages}{032213} (\bibinfo{year}{2019}),
  \urlprefix\url{https://link.aps.org/doi/10.1103/PhysRevE.99.032213}.

\bibitem[{\citenamefont{Kumari and Ghose}(2019)}]{Meenu-2019}
\bibinfo{author}{\bibfnamefont{M.}~\bibnamefont{Kumari}} \bibnamefont{and}
  \bibinfo{author}{\bibfnamefont{S.}~\bibnamefont{Ghose}},
  \bibinfo{journal}{Phys. Rev. A} \textbf{\bibinfo{volume}{99}},
  \bibinfo{pages}{042311} (\bibinfo{year}{2019}),
  \urlprefix\url{https://link.aps.org/doi/10.1103/PhysRevA.99.042311}.

\bibitem[{\citenamefont{Lerose and Pappalardi}(2020)}]{Pappalardi-2020}
\bibinfo{author}{\bibfnamefont{A.}~\bibnamefont{Lerose}} \bibnamefont{and}
  \bibinfo{author}{\bibfnamefont{S.}~\bibnamefont{Pappalardi}},
  \bibinfo{journal}{Phys. Rev. A} \textbf{\bibinfo{volume}{102}},
  \bibinfo{pages}{032404} (\bibinfo{year}{2020}),
  \urlprefix\url{https://link.aps.org/doi/10.1103/PhysRevA.102.032404}.

\bibitem[{\citenamefont{Chaudhury et~al.}(2009)\citenamefont{Chaudhury, Smith,
  Anderson, Ghose, and Jessen}}]{Chaudhary}
\bibinfo{author}{\bibfnamefont{S.}~\bibnamefont{Chaudhury}},
  \bibinfo{author}{\bibfnamefont{A.}~\bibnamefont{Smith}},
  \bibinfo{author}{\bibfnamefont{B.~E.} \bibnamefont{Anderson}},
  \bibinfo{author}{\bibfnamefont{S.}~\bibnamefont{Ghose}}, \bibnamefont{and}
  \bibinfo{author}{\bibfnamefont{P.~S.} \bibnamefont{Jessen}},
  \bibinfo{journal}{Nature} \textbf{\bibinfo{volume}{461}},
  \bibinfo{pages}{768} (\bibinfo{year}{2009}).

\bibitem[{\citenamefont{Haake}(1991)}]{Haake}
\bibinfo{author}{\bibfnamefont{F.}~\bibnamefont{Haake}},
  \emph{\bibinfo{title}{Quantum Signatures of Chaos}}
  (\bibinfo{publisher}{Spring-Verlag, Berlin}, \bibinfo{year}{1991}).

\bibitem[{\citenamefont{Dogra et~al.}(2019)\citenamefont{Dogra, Madhok, and
  Lakshminarayan}}]{dogra19}
\bibinfo{author}{\bibfnamefont{S.}~\bibnamefont{Dogra}},
  \bibinfo{author}{\bibfnamefont{V.}~\bibnamefont{Madhok}}, \bibnamefont{and}
  \bibinfo{author}{\bibfnamefont{A.}~\bibnamefont{Lakshminarayan}},
  \bibinfo{journal}{Phys. Rev. E} \textbf{\bibinfo{volume}{99}},
  \bibinfo{pages}{062217} (\bibinfo{year}{2019}),
  \urlprefix\url{https://link.aps.org/doi/10.1103/PhysRevE.99.062217}.

\bibitem[{\citenamefont{Ruebeck et~al.}(2017)\citenamefont{Ruebeck, Lin, and
  Pattanayak}}]{RuebeckArjendu2017}
\bibinfo{author}{\bibfnamefont{J.~B.} \bibnamefont{Ruebeck}},
  \bibinfo{author}{\bibfnamefont{J.}~\bibnamefont{Lin}}, \bibnamefont{and}
  \bibinfo{author}{\bibfnamefont{A.~K.} \bibnamefont{Pattanayak}},
  \bibinfo{journal}{Phys. Rev. E} \textbf{\bibinfo{volume}{95}},
  \bibinfo{pages}{062222} (\bibinfo{year}{2017}).

\bibitem[{\citenamefont{Bhosale and Santhanam}(2018)}]{Bhosale-2018}
\bibinfo{author}{\bibfnamefont{U.~T.} \bibnamefont{Bhosale}} \bibnamefont{and}
  \bibinfo{author}{\bibfnamefont{M.~S.} \bibnamefont{Santhanam}},
  \bibinfo{journal}{Phys. Rev. E} \textbf{\bibinfo{volume}{98}},
  \bibinfo{pages}{052228} (\bibinfo{year}{2018}),
  \urlprefix\url{https://link.aps.org/doi/10.1103/PhysRevE.98.052228}.

\bibitem[{\citenamefont{Krithika et~al.}(2019)\citenamefont{Krithika, Anjusha,
  Bhosale, and Mahesh}}]{MaheshUdayExpt-2019}
\bibinfo{author}{\bibfnamefont{V.~R.} \bibnamefont{Krithika}},
  \bibinfo{author}{\bibfnamefont{V.~S.} \bibnamefont{Anjusha}},
  \bibinfo{author}{\bibfnamefont{U.~T.} \bibnamefont{Bhosale}},
  \bibnamefont{and} \bibinfo{author}{\bibfnamefont{T.~S.}
  \bibnamefont{Mahesh}}, \bibinfo{journal}{Phys. Rev. E}
  \textbf{\bibinfo{volume}{99}}, \bibinfo{pages}{032219}
  (\bibinfo{year}{2019}),
  \urlprefix\url{https://link.aps.org/doi/10.1103/PhysRevE.99.032219}.

\bibitem[{\citenamefont{Peres}(1984{\natexlab{a}})}]{per00}
\bibinfo{author}{\bibfnamefont{A.}~\bibnamefont{Peres}},
  \bibinfo{journal}{Phys. Rev. A} \textbf{\bibinfo{volume}{30}},
  \bibinfo{pages}{1610} (\bibinfo{year}{1984}{\natexlab{a}}).

\bibitem[{\citenamefont{Schack and Caves}(1996)}]{sc96}
\bibinfo{author}{\bibfnamefont{R.}~\bibnamefont{Schack}} \bibnamefont{and}
  \bibinfo{author}{\bibfnamefont{C.~M.} \bibnamefont{Caves}},
  \bibinfo{journal}{Phys. Rev. E} \textbf{\bibinfo{volume}{53}},
  \bibinfo{pages}{3257} (\bibinfo{year}{1996}).

\bibitem[{\citenamefont{Berry and Tabor}(1977)}]{Berry375}
\bibinfo{author}{\bibfnamefont{M.}~\bibnamefont{Berry}} \bibnamefont{and}
  \bibinfo{author}{\bibfnamefont{M.}~\bibnamefont{Tabor}},
  \bibinfo{journal}{Proceedings of the Royal Society of London A: Mathematical,
  Physical and Engineering Sciences} \textbf{\bibinfo{volume}{356}},
  \bibinfo{pages}{375} (\bibinfo{year}{1977}), ISSN \bibinfo{issn}{0080-4630}.

\bibitem[{\citenamefont{Bohigas and Flores}(1971)}]{bohigas1971spacing}
\bibinfo{author}{\bibfnamefont{O.}~\bibnamefont{Bohigas}} \bibnamefont{and}
  \bibinfo{author}{\bibfnamefont{J.}~\bibnamefont{Flores}},
  \bibinfo{journal}{Physics Letters B} \textbf{\bibinfo{volume}{35}},
  \bibinfo{pages}{383} (\bibinfo{year}{1971}).

\bibitem[{\citenamefont{Berry}(1977)}]{Berry77a}
\bibinfo{author}{\bibfnamefont{M.~V.} \bibnamefont{Berry}},
  \bibinfo{journal}{Journal of Physics A: Mathematical and General}
  \textbf{\bibinfo{volume}{10}}, \bibinfo{pages}{2083} (\bibinfo{year}{1977}),
  \urlprefix\url{http://stacks.iop.org/0305-4470/10/i=12/a=016}.

\bibitem[{\citenamefont{Berry et~al.}(1979)\citenamefont{Berry, Balazs, Tabor,
  and Voros}}]{BERRY197926}
\bibinfo{author}{\bibfnamefont{M.}~\bibnamefont{Berry}},
  \bibinfo{author}{\bibfnamefont{N.}~\bibnamefont{Balazs}},
  \bibinfo{author}{\bibfnamefont{M.}~\bibnamefont{Tabor}}, \bibnamefont{and}
  \bibinfo{author}{\bibfnamefont{A.}~\bibnamefont{Voros}},
  \bibinfo{journal}{Annals of Physics} \textbf{\bibinfo{volume}{122}},
  \bibinfo{pages}{26 } (\bibinfo{year}{1979}), ISSN \bibinfo{issn}{0003-4916},
  \urlprefix\url{http://www.sciencedirect.com/science/article/pii/0003491679902963}.

\bibitem[{\citenamefont{Voros}(1976)}]{voros1976semi}
\bibinfo{author}{\bibfnamefont{A.}~\bibnamefont{Voros}}, \bibinfo{journal}{Ann.
  Inst. Henri Poincar{\'e} A} \textbf{\bibinfo{volume}{24}},
  \bibinfo{pages}{31} (\bibinfo{year}{1976}).

\bibitem[{\citenamefont{Heller}(1984)}]{heller1984bound}
\bibinfo{author}{\bibfnamefont{E.~J.} \bibnamefont{Heller}},
  \bibinfo{journal}{Physical Review Letters} \textbf{\bibinfo{volume}{53}},
  \bibinfo{pages}{1515} (\bibinfo{year}{1984}).

\bibitem[{\citenamefont{McDonald and Kaufman}(1979)}]{McDonald}
\bibinfo{author}{\bibfnamefont{S.~W.} \bibnamefont{McDonald}} \bibnamefont{and}
  \bibinfo{author}{\bibfnamefont{A.~N.} \bibnamefont{Kaufman}},
  \bibinfo{journal}{Phys. Rev. Lett.} \textbf{\bibinfo{volume}{42}},
  \bibinfo{pages}{1189} (\bibinfo{year}{1979}),
  \urlprefix\url{https://link.aps.org/doi/10.1103/PhysRevLett.42.1189}.

\bibitem[{\citenamefont{Robnik and Berry}(1985)}]{robnik1985classical}
\bibinfo{author}{\bibfnamefont{M.}~\bibnamefont{Robnik}} \bibnamefont{and}
  \bibinfo{author}{\bibfnamefont{M.~V.} \bibnamefont{Berry}},
  \bibinfo{journal}{Journal of Physics A: Mathematical and General}
  \textbf{\bibinfo{volume}{18}}, \bibinfo{pages}{1361} (\bibinfo{year}{1985}).

\bibitem[{\citenamefont{Gutzwiller}(1971)}]{gutzwiller1971periodic}
\bibinfo{author}{\bibfnamefont{M.~C.} \bibnamefont{Gutzwiller}},
  \bibinfo{journal}{Journal of Mathematical Physics}
  \textbf{\bibinfo{volume}{12}}, \bibinfo{pages}{343} (\bibinfo{year}{1971}).

\bibitem[{\citenamefont{Swingle et~al.}(2016)\citenamefont{Swingle, Bentsen,
  Schleier-Smith, and Hayden}}]{swingle2016measuring}
\bibinfo{author}{\bibfnamefont{B.}~\bibnamefont{Swingle}},
  \bibinfo{author}{\bibfnamefont{G.}~\bibnamefont{Bentsen}},
  \bibinfo{author}{\bibfnamefont{M.}~\bibnamefont{Schleier-Smith}},
  \bibnamefont{and} \bibinfo{author}{\bibfnamefont{P.}~\bibnamefont{Hayden}},
  \bibinfo{journal}{Physical Review A} \textbf{\bibinfo{volume}{94}},
  \bibinfo{pages}{040302} (\bibinfo{year}{2016}).

\bibitem[{\citenamefont{Hayden and Preskill}(2007)}]{hayden2007black}
\bibinfo{author}{\bibfnamefont{P.}~\bibnamefont{Hayden}} \bibnamefont{and}
  \bibinfo{author}{\bibfnamefont{J.}~\bibnamefont{Preskill}},
  \bibinfo{journal}{Journal of High Energy Physics}
  \textbf{\bibinfo{volume}{2007}}, \bibinfo{pages}{120} (\bibinfo{year}{2007}).

\bibitem[{\citenamefont{Maldacena and Stanford}(2016)}]{MaldacenaSYK15}
\bibinfo{author}{\bibfnamefont{J.}~\bibnamefont{Maldacena}} \bibnamefont{and}
  \bibinfo{author}{\bibfnamefont{D.}~\bibnamefont{Stanford}},
  \bibinfo{journal}{Phys. Rev. D} \textbf{\bibinfo{volume}{94}},
  \bibinfo{pages}{106002} (\bibinfo{year}{2016}),
  \urlprefix\url{https://link.aps.org/doi/10.1103/PhysRevD.94.106002}.

\bibitem[{\citenamefont{Maldacena et~al.}(2016)\citenamefont{Maldacena,
  Shenker, and Stanford}}]{Maldacena2016}
\bibinfo{author}{\bibfnamefont{J.}~\bibnamefont{Maldacena}},
  \bibinfo{author}{\bibfnamefont{S.~H.} \bibnamefont{Shenker}},
  \bibnamefont{and} \bibinfo{author}{\bibfnamefont{D.}~\bibnamefont{Stanford}},
  \bibinfo{journal}{Journal of High Energy Physics}
  \textbf{\bibinfo{volume}{2016}}, \bibinfo{pages}{106} (\bibinfo{year}{2016}),
  ISSN \bibinfo{issn}{1029-8479},
  \urlprefix\url{https://doi.org/10.1007/JHEP08(2016)106}.

\bibitem[{\citenamefont{Hartman and Maldacena}(2013)}]{hartman2013time}
\bibinfo{author}{\bibfnamefont{T.}~\bibnamefont{Hartman}} \bibnamefont{and}
  \bibinfo{author}{\bibfnamefont{J.}~\bibnamefont{Maldacena}},
  \bibinfo{journal}{Journal of High Energy Physics}
  \textbf{\bibinfo{volume}{2013}}, \bibinfo{pages}{14} (\bibinfo{year}{2013}).

\bibitem[{\citenamefont{Shenker and Stanford}(2014)}]{shenker2014black}
\bibinfo{author}{\bibfnamefont{S.~H.} \bibnamefont{Shenker}} \bibnamefont{and}
  \bibinfo{author}{\bibfnamefont{D.}~\bibnamefont{Stanford}},
  \bibinfo{journal}{Journal of High Energy Physics}
  \textbf{\bibinfo{volume}{2014}}, \bibinfo{pages}{67} (\bibinfo{year}{2014}).

\bibitem[{\citenamefont{Sekino and Susskind}(2008)}]{sekino2008fast}
\bibinfo{author}{\bibfnamefont{Y.}~\bibnamefont{Sekino}} \bibnamefont{and}
  \bibinfo{author}{\bibfnamefont{L.}~\bibnamefont{Susskind}},
  \bibinfo{journal}{Journal of High Energy Physics}
  \textbf{\bibinfo{volume}{2008}}, \bibinfo{pages}{065} (\bibinfo{year}{2008}).

\bibitem[{\citenamefont{Lashkari et~al.}(2013)\citenamefont{Lashkari, Stanford,
  Hastings, Osborne, and Hayden}}]{Lashkari2013}
\bibinfo{author}{\bibfnamefont{N.}~\bibnamefont{Lashkari}},
  \bibinfo{author}{\bibfnamefont{D.}~\bibnamefont{Stanford}},
  \bibinfo{author}{\bibfnamefont{M.}~\bibnamefont{Hastings}},
  \bibinfo{author}{\bibfnamefont{T.}~\bibnamefont{Osborne}}, \bibnamefont{and}
  \bibinfo{author}{\bibfnamefont{P.}~\bibnamefont{Hayden}},
  \bibinfo{journal}{Journal of High Energy Physics}
  \textbf{\bibinfo{volume}{2013}}, \bibinfo{pages}{22} (\bibinfo{year}{2013}),
  ISSN \bibinfo{issn}{1029-8479},
  \urlprefix\url{https://doi.org/10.1007/JHEP04(2013)022}.

\bibitem[{\citenamefont{Hosur et~al.}(2016)\citenamefont{Hosur, Qi, Roberts,
  and Yoshida}}]{hosur2016chaos}
\bibinfo{author}{\bibfnamefont{P.}~\bibnamefont{Hosur}},
  \bibinfo{author}{\bibfnamefont{X.-L.} \bibnamefont{Qi}},
  \bibinfo{author}{\bibfnamefont{D.~A.} \bibnamefont{Roberts}},
  \bibnamefont{and} \bibinfo{author}{\bibfnamefont{B.}~\bibnamefont{Yoshida}},
  \bibinfo{journal}{Journal of High Energy Physics}
  \textbf{\bibinfo{volume}{2016}}, \bibinfo{pages}{4} (\bibinfo{year}{2016}).

\bibitem[{\citenamefont{Iyoda and Sagawa}(2017)}]{IyodaSagawa}
\bibinfo{author}{\bibfnamefont{E.}~\bibnamefont{Iyoda}} \bibnamefont{and}
  \bibinfo{author}{\bibfnamefont{T.}~\bibnamefont{Sagawa}},
  \bibinfo{journal}{arXiv:1704.04850}  (\bibinfo{year}{2017}).

\bibitem[{\citenamefont{Larkin and Ovchinnikov}(1969)}]{Larkin-1969}
\bibinfo{author}{\bibfnamefont{A.~I.} \bibnamefont{Larkin}} \bibnamefont{and}
  \bibinfo{author}{\bibfnamefont{Y.~N.} \bibnamefont{Ovchinnikov}},
  \bibinfo{journal}{JETP} \textbf{\bibinfo{volume}{28}}, \bibinfo{pages}{1200}
  (\bibinfo{year}{1969}).

\bibitem[{\citenamefont{Cotler et~al.}(2018)\citenamefont{Cotler, Ding, and
  Penington}}]{Cotler-2018}
\bibinfo{author}{\bibfnamefont{J.~S.} \bibnamefont{Cotler}},
  \bibinfo{author}{\bibfnamefont{D.}~\bibnamefont{Ding}}, \bibnamefont{and}
  \bibinfo{author}{\bibfnamefont{G.~R.} \bibnamefont{Penington}},
  \bibinfo{journal}{Ann. Phys.} \textbf{\bibinfo{volume}{396}},
  \bibinfo{pages}{318 } (\bibinfo{year}{2018}), ISSN \bibinfo{issn}{0003-4916},
  \urlprefix\url{http://www.sciencedirect.com/science/article/pii/S000349161830191X}.

\bibitem[{\citenamefont{Hashimoto et~al.}(2017)\citenamefont{Hashimoto, Murata,
  and Yoshii}}]{Hashimoto-2017}
\bibinfo{author}{\bibfnamefont{K.}~\bibnamefont{Hashimoto}},
  \bibinfo{author}{\bibfnamefont{K.}~\bibnamefont{Murata}}, \bibnamefont{and}
  \bibinfo{author}{\bibfnamefont{R.}~\bibnamefont{Yoshii}},
  \bibinfo{journal}{J. High Energy Phys.} \textbf{\bibinfo{volume}{2017}},
  \bibinfo{pages}{138} (\bibinfo{year}{2017}), ISSN \bibinfo{issn}{1029-8479}.

\bibitem[{\citenamefont{Swingle}(2018)}]{Swingle-2018}
\bibinfo{author}{\bibfnamefont{B.}~\bibnamefont{Swingle}},
  \bibinfo{journal}{Nature Physics} \textbf{\bibinfo{volume}{14}},
  \bibinfo{pages}{988} (\bibinfo{year}{2018}).

\bibitem[{\citenamefont{Ch\'avez-Carlos
  et~al.}(2019)\citenamefont{Ch\'avez-Carlos, L\'opez-del Carpio,
  Bastarrachea-Magnani, Str\'ansk\'y, Lerma-Hern\'andez, Santos, and
  Hirsch}}]{Carlos-2019}
\bibinfo{author}{\bibfnamefont{J.}~\bibnamefont{Ch\'avez-Carlos}},
  \bibinfo{author}{\bibfnamefont{B.}~\bibnamefont{L\'opez-del Carpio}},
  \bibinfo{author}{\bibfnamefont{M.~A.} \bibnamefont{Bastarrachea-Magnani}},
  \bibinfo{author}{\bibfnamefont{P.}~\bibnamefont{Str\'ansk\'y}},
  \bibinfo{author}{\bibfnamefont{S.}~\bibnamefont{Lerma-Hern\'andez}},
  \bibinfo{author}{\bibfnamefont{L.~F.} \bibnamefont{Santos}},
  \bibnamefont{and} \bibinfo{author}{\bibfnamefont{J.~G.}
  \bibnamefont{Hirsch}}, \bibinfo{journal}{Phys. Rev. Lett.}
  \textbf{\bibinfo{volume}{122}}, \bibinfo{pages}{024101}
  (\bibinfo{year}{2019}),
  \urlprefix\url{https://link.aps.org/doi/10.1103/PhysRevLett.122.024101}.

\bibitem[{\citenamefont{Rozenbaum et~al.}(2017)\citenamefont{Rozenbaum,
  Ganeshan, and Galitski}}]{Rozenbaum17}
\bibinfo{author}{\bibfnamefont{E.~B.} \bibnamefont{Rozenbaum}},
  \bibinfo{author}{\bibfnamefont{S.}~\bibnamefont{Ganeshan}}, \bibnamefont{and}
  \bibinfo{author}{\bibfnamefont{V.}~\bibnamefont{Galitski}},
  \bibinfo{journal}{Phys. Rev. Lett.} \textbf{\bibinfo{volume}{118}},
  \bibinfo{pages}{086801} (\bibinfo{year}{2017}),
  \urlprefix\url{https://link.aps.org/doi/10.1103/PhysRevLett.118.086801}.

\bibitem[{\citenamefont{Lakshminarayan}(2019)}]{Arul-Baker-2018}
\bibinfo{author}{\bibfnamefont{A.}~\bibnamefont{Lakshminarayan}},
  \bibinfo{journal}{Phys. Rev. E} \textbf{\bibinfo{volume}{99}},
  \bibinfo{pages}{012201} (\bibinfo{year}{2019}),
  \urlprefix\url{https://link.aps.org/doi/10.1103/PhysRevE.99.012201}.

\bibitem[{\citenamefont{Garc\'{\i}a-Mata
  et~al.}(2018)\citenamefont{Garc\'{\i}a-Mata, Saraceno, Jalabert, Roncaglia,
  and Wisniacki}}]{Saraceno-2018}
\bibinfo{author}{\bibfnamefont{I.}~\bibnamefont{Garc\'{\i}a-Mata}},
  \bibinfo{author}{\bibfnamefont{M.}~\bibnamefont{Saraceno}},
  \bibinfo{author}{\bibfnamefont{R.~A.} \bibnamefont{Jalabert}},
  \bibinfo{author}{\bibfnamefont{A.~J.} \bibnamefont{Roncaglia}},
  \bibnamefont{and} \bibinfo{author}{\bibfnamefont{D.~A.}
  \bibnamefont{Wisniacki}}, \bibinfo{journal}{Phys. Rev. Lett.}
  \textbf{\bibinfo{volume}{121}}, \bibinfo{pages}{210601}
  (\bibinfo{year}{2018}),
  \urlprefix\url{https://link.aps.org/doi/10.1103/PhysRevLett.121.210601}.

\bibitem[{\citenamefont{Moudgalya et~al.}(2019)\citenamefont{Moudgalya,
  Devakul, von Keyserlingk, and Sondhi}}]{Moudgalya-2018}
\bibinfo{author}{\bibfnamefont{S.}~\bibnamefont{Moudgalya}},
  \bibinfo{author}{\bibfnamefont{T.}~\bibnamefont{Devakul}},
  \bibinfo{author}{\bibfnamefont{C.~W.} \bibnamefont{von Keyserlingk}},
  \bibnamefont{and} \bibinfo{author}{\bibfnamefont{S.~L.}
  \bibnamefont{Sondhi}}, \bibinfo{journal}{Phys. Rev. B}
  \textbf{\bibinfo{volume}{99}}, \bibinfo{pages}{094312}
  (\bibinfo{year}{2019}),
  \urlprefix\url{https://link.aps.org/doi/10.1103/PhysRevB.99.094312}.

\bibitem[{\citenamefont{Rammensee et~al.}(2018)\citenamefont{Rammensee, Urbina,
  and Richter}}]{Klaus-2018}
\bibinfo{author}{\bibfnamefont{J.}~\bibnamefont{Rammensee}},
  \bibinfo{author}{\bibfnamefont{J.~D.} \bibnamefont{Urbina}},
  \bibnamefont{and} \bibinfo{author}{\bibfnamefont{K.}~\bibnamefont{Richter}},
  \bibinfo{journal}{Phys. Rev. Lett.} \textbf{\bibinfo{volume}{121}},
  \bibinfo{pages}{124101} (\bibinfo{year}{2018}),
  \urlprefix\url{https://link.aps.org/doi/10.1103/PhysRevLett.121.124101}.

\bibitem[{\citenamefont{Jalabert et~al.}(2018)\citenamefont{Jalabert,
  Garc\'{\i}a-Mata, and Wisniacki}}]{Jalabert-2018}
\bibinfo{author}{\bibfnamefont{R.~A.} \bibnamefont{Jalabert}},
  \bibinfo{author}{\bibfnamefont{I.}~\bibnamefont{Garc\'{\i}a-Mata}},
  \bibnamefont{and} \bibinfo{author}{\bibfnamefont{D.~A.}
  \bibnamefont{Wisniacki}}, \bibinfo{journal}{Phys. Rev. E}
  \textbf{\bibinfo{volume}{98}}, \bibinfo{pages}{062218}
  (\bibinfo{year}{2018}),
  \urlprefix\url{https://link.aps.org/doi/10.1103/PhysRevE.98.062218}.

\bibitem[{\citenamefont{Prakash and Lakshminarayan}(2020)}]{Ravi-Arul-2020}
\bibinfo{author}{\bibfnamefont{R.}~\bibnamefont{Prakash}} \bibnamefont{and}
  \bibinfo{author}{\bibfnamefont{A.}~\bibnamefont{Lakshminarayan}},
  \bibinfo{journal}{Phys. Rev. B} \textbf{\bibinfo{volume}{101}},
  \bibinfo{pages}{121108} (\bibinfo{year}{2020}),
  \urlprefix\url{https://link.aps.org/doi/10.1103/PhysRevB.101.121108}.

\bibitem[{\citenamefont{Seshadri et~al.}(2018)\citenamefont{Seshadri, Madhok,
  and Lakshminarayan}}]{Akshay2018}
\bibinfo{author}{\bibfnamefont{A.}~\bibnamefont{Seshadri}},
  \bibinfo{author}{\bibfnamefont{V.}~\bibnamefont{Madhok}}, \bibnamefont{and}
  \bibinfo{author}{\bibfnamefont{A.}~\bibnamefont{Lakshminarayan}},
  \bibinfo{journal}{Phys. Rev. E} \textbf{\bibinfo{volume}{98}},
  \bibinfo{pages}{052205} (\bibinfo{year}{2018}),
  \urlprefix\url{https://link.aps.org/doi/10.1103/PhysRevE.98.052205}.

\bibitem[{\citenamefont{Sieberer et~al.}(2019)\citenamefont{Sieberer, Olsacher,
  Elben, Heyl, Hauke, Haake, and Zoller}}]{sieberer2019digital}
\bibinfo{author}{\bibfnamefont{L.~M.} \bibnamefont{Sieberer}},
  \bibinfo{author}{\bibfnamefont{T.}~\bibnamefont{Olsacher}},
  \bibinfo{author}{\bibfnamefont{A.}~\bibnamefont{Elben}},
  \bibinfo{author}{\bibfnamefont{M.}~\bibnamefont{Heyl}},
  \bibinfo{author}{\bibfnamefont{P.}~\bibnamefont{Hauke}},
  \bibinfo{author}{\bibfnamefont{F.}~\bibnamefont{Haake}}, \bibnamefont{and}
  \bibinfo{author}{\bibfnamefont{P.}~\bibnamefont{Zoller}},
  \bibinfo{journal}{npj Quantum Information} \textbf{\bibinfo{volume}{5}},
  \bibinfo{pages}{1} (\bibinfo{year}{2019}).

\bibitem[{\citenamefont{Yin and Lucas}(2020)}]{yin2020quantum}
\bibinfo{author}{\bibfnamefont{C.}~\bibnamefont{Yin}} \bibnamefont{and}
  \bibinfo{author}{\bibfnamefont{A.}~\bibnamefont{Lucas}},
  \emph{\bibinfo{title}{Quantum operator growth bounds for kicked tops and
  semiclassical spin chains}} (\bibinfo{year}{2020}), \eprint{2010.06592}.

\bibitem[{\citenamefont{Peres}(2006)}]{peres2006quantum}
\bibinfo{author}{\bibfnamefont{A.}~\bibnamefont{Peres}},
  \emph{\bibinfo{title}{Quantum theory: concepts and methods}},
  vol.~\bibinfo{volume}{57} (\bibinfo{publisher}{Springer Science \& Business
  Media}, \bibinfo{year}{2006}).

\bibitem[{\citenamefont{Peres}(1984{\natexlab{b}})}]{peres1984stability}
\bibinfo{author}{\bibfnamefont{A.}~\bibnamefont{Peres}},
  \bibinfo{journal}{Physical Review A} \textbf{\bibinfo{volume}{30}},
  \bibinfo{pages}{1610} (\bibinfo{year}{1984}{\natexlab{b}}).

\bibitem[{\citenamefont{Prosen et~al.}(2003)\citenamefont{Prosen, Seligman, and
  {\v{Z}}nidari{\v{c}}}}]{prosen2003theory}
\bibinfo{author}{\bibfnamefont{T.}~\bibnamefont{Prosen}},
  \bibinfo{author}{\bibfnamefont{T.~H.} \bibnamefont{Seligman}},
  \bibnamefont{and}
  \bibinfo{author}{\bibfnamefont{M.}~\bibnamefont{{\v{Z}}nidari{\v{c}}}},
  \bibinfo{journal}{Progress of Theoretical Physics Supplement}
  \textbf{\bibinfo{volume}{150}}, \bibinfo{pages}{200} (\bibinfo{year}{2003}).

\bibitem[{\citenamefont{Gorin et~al.}(2006)\citenamefont{Gorin, Prosen,
  Seligman, and {\v{Z}}nidari{\v{c}}}}]{gorin2006dynamics}
\bibinfo{author}{\bibfnamefont{T.}~\bibnamefont{Gorin}},
  \bibinfo{author}{\bibfnamefont{T.}~\bibnamefont{Prosen}},
  \bibinfo{author}{\bibfnamefont{T.~H.} \bibnamefont{Seligman}},
  \bibnamefont{and}
  \bibinfo{author}{\bibfnamefont{M.}~\bibnamefont{{\v{Z}}nidari{\v{c}}}},
  \bibinfo{journal}{Physics Reports} \textbf{\bibinfo{volume}{435}},
  \bibinfo{pages}{33} (\bibinfo{year}{2006}).

\bibitem[{\citenamefont{Sankaranarayanan and
  Lakshminarayan}(2003)}]{PhysRevE.68.036216}
\bibinfo{author}{\bibfnamefont{R.}~\bibnamefont{Sankaranarayanan}}
  \bibnamefont{and}
  \bibinfo{author}{\bibfnamefont{A.}~\bibnamefont{Lakshminarayan}},
  \bibinfo{journal}{Phys. Rev. E} \textbf{\bibinfo{volume}{68}},
  \bibinfo{pages}{036216} (\bibinfo{year}{2003}),
  \urlprefix\url{https://link.aps.org/doi/10.1103/PhysRevE.68.036216}.

\bibitem[{\citenamefont{Zurek and Paz}(1994)}]{Zurek/Paz}
\bibinfo{author}{\bibfnamefont{W.~H.} \bibnamefont{Zurek}} \bibnamefont{and}
  \bibinfo{author}{\bibfnamefont{J.~P.} \bibnamefont{Paz}},
  \bibinfo{journal}{Phys. Rev. Lett.} \textbf{\bibinfo{volume}{72}},
  \bibinfo{pages}{2508} (\bibinfo{year}{1994}).

\bibitem[{\citenamefont{Georgeot and
  Shepelyansky}(2000{\natexlab{a}})}]{PhysRevE.62.3504}
\bibinfo{author}{\bibfnamefont{B.}~\bibnamefont{Georgeot}} \bibnamefont{and}
  \bibinfo{author}{\bibfnamefont{D.~L.} \bibnamefont{Shepelyansky}},
  \bibinfo{journal}{Phys. Rev. E} \textbf{\bibinfo{volume}{62}},
  \bibinfo{pages}{3504} (\bibinfo{year}{2000}{\natexlab{a}}),
  \urlprefix\url{https://link.aps.org/doi/10.1103/PhysRevE.62.3504}.

\bibitem[{\citenamefont{Georgeot and
  Shepelyansky}(2000{\natexlab{b}})}]{PhysRevE.62.6366}
\bibinfo{author}{\bibfnamefont{B.}~\bibnamefont{Georgeot}} \bibnamefont{and}
  \bibinfo{author}{\bibfnamefont{D.~L.} \bibnamefont{Shepelyansky}},
  \bibinfo{journal}{Phys. Rev. E} \textbf{\bibinfo{volume}{62}},
  \bibinfo{pages}{6366} (\bibinfo{year}{2000}{\natexlab{b}}),
  \urlprefix\url{https://link.aps.org/doi/10.1103/PhysRevE.62.6366}.

\bibitem[{\citenamefont{Hauke et~al.}(2012)\citenamefont{Hauke, Cucchietti,
  Tagliacozzo, Deutsch, and Lewenstein}}]{hauke2012can}
\bibinfo{author}{\bibfnamefont{P.}~\bibnamefont{Hauke}},
  \bibinfo{author}{\bibfnamefont{F.~M.} \bibnamefont{Cucchietti}},
  \bibinfo{author}{\bibfnamefont{L.}~\bibnamefont{Tagliacozzo}},
  \bibinfo{author}{\bibfnamefont{I.}~\bibnamefont{Deutsch}}, \bibnamefont{and}
  \bibinfo{author}{\bibfnamefont{M.}~\bibnamefont{Lewenstein}},
  \bibinfo{journal}{Reports on Progress in Physics}
  \textbf{\bibinfo{volume}{75}}, \bibinfo{pages}{082401}
  (\bibinfo{year}{2012}).

\bibitem[{\citenamefont{Georgescu et~al.}(2014)\citenamefont{Georgescu, Ashhab,
  and Nori}}]{georgescu2014quantum}
\bibinfo{author}{\bibfnamefont{I.~M.} \bibnamefont{Georgescu}},
  \bibinfo{author}{\bibfnamefont{S.}~\bibnamefont{Ashhab}}, \bibnamefont{and}
  \bibinfo{author}{\bibfnamefont{F.}~\bibnamefont{Nori}},
  \bibinfo{journal}{Reviews of Modern Physics} \textbf{\bibinfo{volume}{86}},
  \bibinfo{pages}{153} (\bibinfo{year}{2014}).

\bibitem[{\citenamefont{Deutsch}(2020)}]{deutsch2020harnessing}
\bibinfo{author}{\bibfnamefont{I.~H.} \bibnamefont{Deutsch}},
  \bibinfo{journal}{arXiv preprint arXiv:2010.10283}  (\bibinfo{year}{2020}).

\bibitem[{\citenamefont{Fiderer and Braun}(2018)}]{fiderer2018quantum}
\bibinfo{author}{\bibfnamefont{L.~J.} \bibnamefont{Fiderer}} \bibnamefont{and}
  \bibinfo{author}{\bibfnamefont{D.}~\bibnamefont{Braun}},
  \bibinfo{journal}{Nature communications} \textbf{\bibinfo{volume}{9}},
  \bibinfo{pages}{1} (\bibinfo{year}{2018}).

\bibitem[{\citenamefont{Ku{\'s} et~al.}(1987)\citenamefont{Ku{\'s}, Scharf, and
  Haake}}]{KusScharfHaake1987}
\bibinfo{author}{\bibfnamefont{M.}~\bibnamefont{Ku{\'s}}},
  \bibinfo{author}{\bibfnamefont{R.}~\bibnamefont{Scharf}}, \bibnamefont{and}
  \bibinfo{author}{\bibfnamefont{F.}~\bibnamefont{Haake}},
  \bibinfo{journal}{Zeitschrift f{\"u}r Physik B Condensed Matter}
  \textbf{\bibinfo{volume}{66}}, \bibinfo{pages}{129} (\bibinfo{year}{1987}).

\bibitem[{\citenamefont{Peres}(2002)}]{Peres02}
\bibinfo{author}{\bibfnamefont{A.}~\bibnamefont{Peres}},
  \emph{\bibinfo{title}{Quantum Theory: Concepts and Methods}}
  (\bibinfo{publisher}{Kluwer Academic Publishers}, \bibinfo{address}{New
  York}, \bibinfo{year}{2002}).

\bibitem[{\citenamefont{Milburn}(1999)}]{Milburn99}
\bibinfo{author}{\bibfnamefont{G.~J.} \bibnamefont{Milburn}},
  \emph{\bibinfo{title}{Simulating nonlinear spin models in an ion trap}}
  (\bibinfo{year}{1999}), \eprint{arXiv:quant-ph/9908037}.

\bibitem[{\citenamefont{Wang et~al.}(2004)\citenamefont{Wang, Ghose, Sanders,
  and Hu}}]{Wang2004}
\bibinfo{author}{\bibfnamefont{X.}~\bibnamefont{Wang}},
  \bibinfo{author}{\bibfnamefont{S.}~\bibnamefont{Ghose}},
  \bibinfo{author}{\bibfnamefont{B.~C.} \bibnamefont{Sanders}},
  \bibnamefont{and} \bibinfo{author}{\bibfnamefont{B.}~\bibnamefont{Hu}},
  \bibinfo{journal}{Phys. Rev. E} \textbf{\bibinfo{volume}{70}},
  \bibinfo{pages}{016217} (\bibinfo{year}{2004}).

\bibitem[{\citenamefont{Prosen}(2000)}]{Prosen2000}
\bibinfo{author}{\bibfnamefont{T.}~\bibnamefont{Prosen}},
  \bibinfo{journal}{Progress of Theoretical Physics Supplement}
  \textbf{\bibinfo{volume}{139}}, \bibinfo{pages}{191} (\bibinfo{year}{2000}).

\bibitem[{\citenamefont{{Yan} et~al.}(2019)\citenamefont{{Yan}, {Cincio}, and
  {Zurek}}}]{Yan-2019}
\bibinfo{author}{\bibfnamefont{B.}~\bibnamefont{{Yan}}},
  \bibinfo{author}{\bibfnamefont{L.}~\bibnamefont{{Cincio}}}, \bibnamefont{and}
  \bibinfo{author}{\bibfnamefont{W.~H.} \bibnamefont{{Zurek}}},
  \bibinfo{journal}{arXiv e-prints} \bibinfo{eid}{arXiv:1903.02651}
  (\bibinfo{year}{2019}), \eprint{1903.02651}.

\bibitem[{\citenamefont{Omanakuttan and Lakshminarayan}(2019)}]{SivaAL-2019}
\bibinfo{author}{\bibfnamefont{S.}~\bibnamefont{Omanakuttan}} \bibnamefont{and}
  \bibinfo{author}{\bibfnamefont{A.}~\bibnamefont{Lakshminarayan}},
  \bibinfo{journal}{Phys. Rev. E} \textbf{\bibinfo{volume}{99}},
  \bibinfo{pages}{062128} (\bibinfo{year}{2019}),
  \urlprefix\url{https://link.aps.org/doi/10.1103/PhysRevE.99.062128}.

\bibitem[{\citenamefont{Prakash and Lakshminarayan}(2019)}]{Prakash-2019}
\bibinfo{author}{\bibfnamefont{R.}~\bibnamefont{Prakash}} \bibnamefont{and}
  \bibinfo{author}{\bibfnamefont{A.}~\bibnamefont{Lakshminarayan}},
  \bibinfo{journal}{Acta Phys. Polon. A} \textbf{\bibinfo{volume}{136}},
  \bibinfo{pages}{803} (\bibinfo{year}{2019}), \eprint{1911.02829}.

\bibitem[{\citenamefont{Garc{\'\i}a-Mata
  et~al.}(2016)\citenamefont{Garc{\'\i}a-Mata, Roncaglia, and
  Wisniacki}}]{garcia2016lyapunov}
\bibinfo{author}{\bibfnamefont{I.}~\bibnamefont{Garc{\'\i}a-Mata}},
  \bibinfo{author}{\bibfnamefont{A.~J.} \bibnamefont{Roncaglia}},
  \bibnamefont{and} \bibinfo{author}{\bibfnamefont{D.~A.}
  \bibnamefont{Wisniacki}}, \bibinfo{journal}{Philosophical Transactions of the
  Royal Society A: Mathematical, Physical and Engineering Sciences}
  \textbf{\bibinfo{volume}{374}}, \bibinfo{pages}{20150157}
  (\bibinfo{year}{2016}).

\bibitem[{\citenamefont{Zanardi and Lidar}(2004)}]{zanardi2004purity}
\bibinfo{author}{\bibfnamefont{P.}~\bibnamefont{Zanardi}} \bibnamefont{and}
  \bibinfo{author}{\bibfnamefont{D.~A.} \bibnamefont{Lidar}},
  \bibinfo{journal}{Physical Review A} \textbf{\bibinfo{volume}{70}},
  \bibinfo{pages}{012315} (\bibinfo{year}{2004}).

\bibitem[{\citenamefont{Glauber and Haake}(1976)}]{Glauber}
\bibinfo{author}{\bibfnamefont{R.~J.} \bibnamefont{Glauber}} \bibnamefont{and}
  \bibinfo{author}{\bibfnamefont{F.}~\bibnamefont{Haake}},
  \bibinfo{journal}{Phys. Rev. A} \textbf{\bibinfo{volume}{13}},
  \bibinfo{pages}{357} (\bibinfo{year}{1976}).

\bibitem[{\citenamefont{Puri}(2001)}]{Puri}
\bibinfo{author}{\bibfnamefont{R.~R.} \bibnamefont{Puri}},
  \emph{\bibinfo{title}{Mathematical Methods of Quantum Optics}}
  (\bibinfo{publisher}{Springer}, \bibinfo{address}{Berlin},
  \bibinfo{year}{2001}).

\bibitem[{\citenamefont{Pandey et~al.}(2020)\citenamefont{Pandey, Claeys,
  Campbell, Polkovnikov, and Sels}}]{PhysRevX.10.041017}
\bibinfo{author}{\bibfnamefont{M.}~\bibnamefont{Pandey}},
  \bibinfo{author}{\bibfnamefont{P.~W.} \bibnamefont{Claeys}},
  \bibinfo{author}{\bibfnamefont{D.~K.} \bibnamefont{Campbell}},
  \bibinfo{author}{\bibfnamefont{A.}~\bibnamefont{Polkovnikov}},
  \bibnamefont{and} \bibinfo{author}{\bibfnamefont{D.}~\bibnamefont{Sels}},
  \bibinfo{journal}{Phys. Rev. X} \textbf{\bibinfo{volume}{10}},
  \bibinfo{pages}{041017} (\bibinfo{year}{2020}),
  \urlprefix\url{https://link.aps.org/doi/10.1103/PhysRevX.10.041017}.

\bibitem[{\citenamefont{Trail et~al.}(2008{\natexlab{b}})\citenamefont{Trail,
  Madhok, and Deutsch}}]{tmd08}
\bibinfo{author}{\bibfnamefont{C.~M.} \bibnamefont{Trail}},
  \bibinfo{author}{\bibfnamefont{V.}~\bibnamefont{Madhok}}, \bibnamefont{and}
  \bibinfo{author}{\bibfnamefont{I.~H.} \bibnamefont{Deutsch}},
  \bibinfo{journal}{Phys. Rev. E} \textbf{\bibinfo{volume}{78}},
  \bibinfo{pages}{046211} (\bibinfo{year}{2008}{\natexlab{b}}).

\end{thebibliography}

\end{document}